\documentclass[journal, draftcls, onecolumn]{IEEEtran}
\usepackage[utf8]{inputenc}
\usepackage{amsmath,amssymb}
\usepackage{mathtools}
\usepackage{bbm}

\usepackage{enumitem}

\usepackage{color}

\newtheorem{thm}{Theorem}
\newtheorem{lem}[thm]{Lemma}

\newtheorem{cor}{Corollary}

\newtheorem{rem}[thm]{Remark}

\newtheorem{defn}[thm]{Definition}

\DeclareMathOperator{\supp}{supp}
\newcommand{\ran}{\mathrm{ran}}

\newcommand{\id}{\mathrm{id}}
\newcommand{\mean}{\mathrm{mean}}
\newcommand{\pinv}{{\pi\text{-}\mathrm{inv}}}

\begin{document}

\title{A channel under simultaneous jamming and eavesdropping attack---correlated random coding capacities under strong secrecy criteria}

\author{Moritz Wiese, Janis N\"otzel, Holger Boche
	\thanks{Moritz Wiese has been with the ACCESS Linnaeus Center and the Automatic Control Lab, School of Electrical Engineering, KTH Royal Institute of Technology, Stockholm, Sweden. He is now with the ACCESS Linnaeus Center and the Networked Systems Security Group at KTH Royal Institute of Technology. E-mail: moritzw@kth.se}
	\thanks{Janis N\"otzel has been with the Lehrstuhl f\"ur Theoretische Informationstechnik, Technische Universit\"at M\"unchen, Munich, Germany. He is now with the F\'{\i}sica Te\`{o}rica: Informaci\'{o} i Fen\`{o}mens Qu\`{a}ntics group, Departament de F\'{\i}sica, Universitat Aut\`{o}noma de Barcelona, ES-08193 Bellaterra (Barcelona), Spain. E-mail:Janis.Notzel@uab.cat}
	\thanks{Holger Boche is with the Lehrstuhl f\"ur Theoretische Informationstechnik, Technische Universit\"at M\"unchen, Munich, Germany. E-mail:boche@tum.de}
	\thanks{J.~N\"otzel was supported by the German Research Foundation (DFG) via grant NO 1129/1-1, by the German Federal Ministry of Education and Research via grant 01BQ1050, by the Spanish MINECO Project No. FIS2013-40627-P  as well as the the Generalitat de Catalunya CIRIT Project No. 2014 SGR 966. Holger Boche was supported by the German Federal Ministry of Education and Research via grant 01BQ1050. This paper was presented in part at the  2015 IEEE International Symposium on Information Theory (ISIT '15) in Hong Kong.}
}

\maketitle

\begin{abstract}
	We give a complete characterization of the correlated random coding secrecy capacity of arbitrarily varying wiretap channels (AVWCs). We apply two alternative strong secrecy criteria, which both lead to the same multi-letter formula. The difference of these criteria lies in the treatment of correlated randomness, they coincide in the case of uncorrelated codes. On the basis of the derived formula, we show that the correlated random coding secrecy capacity is continuous as a function of the AVWC, in contrast to the discontinuous uncorrelated coding secrecy capacity. In the proof of the secrecy capacity formula for correlated random codes, we apply an auxiliary channel which is compound from the sender to the intended receiver and arbitrarily varying from the sender to the eavesdropper.
\end{abstract}

\section{Introduction}

This paper brings together two areas of information theory: the arbitrarily varying channel (AVC) and the wiretap channel. This leads to the arbitrarily varying wiretap channel (AVWC): A sender would like to send information to a receiver through a noisy channel. Communication over this channel is subject to two difficulties. First, there is a second receiver, called an eavesdropper, which obtains its own noisy version of the channel inputs and should not be able to decode any information. Second, the state of the channels both to the intended receiver as well as to the eavesdropper can vary arbitrarily over time. Neither the sender nor the intended receiver know the true channel state. For a blocklength $n$, this means that the probability of the intended receiver obtaining the output sequence $y^n=(y_1,\ldots,y_n)$ and the eavesdropper receiving $z^n=(z_1,\ldots,z_n)$ given that $x^n=(x_1,\ldots,x_n)$ was input to the channel is contained in the family
\begin{equation}\label{eq:AVWCfamily}
  \Bigl\{U^n_{s^n}(y^n,z^n\vert x^n)=\prod_{i=1}^nU_{s_i}(y_i,z_i\vert x_i):s^n=(s_1,\ldots,s_n)\in\mathcal S^n\Bigr\}.
\end{equation}
Here, $\mathcal S$ is the finite state set and $\{U_s(\cdot,\cdot\vert\cdot):s\in\mathcal S\}$ a family of stochastic matrices, which thus determines the AVWC.

One could regard the varying channel states as determined by nature. However, we will interpret them as the result of jamming from an intruder. So henceforth, we shall view the AVWC as a channel under two attacks at the same time: one passive (eavesdropping), one active (jamming).

The study of correlated random coding capacities in their own right instead of as mathematical tools applied in the proofs of uncorrelated coding capacity theorems is motivated by arbitarily varying channels (AVCs), which are AVWCs without the eavesdropper. By uncorrelated codes, we mean that sender and receiver have agreed on a procedure $(f,\phi)$ of data manipulation prior to transmission. Here, $f$ is a possibly stochastic mapping from the messages to the channel inputs of a fixed blocklength, $\phi$ reverts channel outputs into messages. For transmission, each node separately executes its part of this procedure without relying on any further resources, in particular no common resources. What we call correlated random coding is usually called random coding and has been used as a mathematical tool ever since Shannon's 1948 paper \cite{S48}. Operationally, it means that sender and receiver agree on a family of deterministic codes $\{(f^\gamma,\phi^\gamma):\gamma\in\Gamma\}$. Before communication, a random experiment following the distribution $\mu$ on $\Gamma$ is performed. The outcome, say $\gamma$, is revealed to sender and intended receiver which then apply the deterministic code $(f^\gamma,\phi^\gamma)$. 

It was already observed by Blackwell, Breiman and Thomasian \cite{BBTa} that whether correlated randomness is available to sender and receiver can be crucial when it comes to the AVC capacity. In fact, AVCs exhibit a dichotomy \cite{A1}: Their capacity for deterministic coding either equals their capacity for correlated random coding or it equals zero. Csisz\'ar and Narayan have identified the distinguishing property \cite{CN}, called symmetrizability (a concept originally introduced by Ericson \cite{E}). Without the use of correlated random coding, a symmetrizable AVC is useless; no message transmission is possible.

Thus one is led to regarding correlated randomness as an additional resource for communication. This resource can make communication possible where it is impossible without. Of course, it is important that the jammer has no access to this resource, i.~e.\ that it does not know the outcome of the random experiment common to sender and receiver. In this paper, we will apply two strong secrecy criteria and show that the corresponding capacities for correlated random coding coincide. The first of these criteria is that
\begin{equation}\label{eq:secBBSa}
  \max_{s^n}\sum_{\gamma}I(M\wedge Z^\gamma_{s^n})\mu(\gamma)
\end{equation}
be small, where $M$ is the message chosen uniformly at random and $Z^\gamma_{s^n}$ is the eavesdropper's output if the state sequence is $s^n$ and the deterministic code $(f^\gamma,\phi^\gamma)$ has been selected. This criterion was applied in \cite{BBSa,MBL}. The second, stronger one requires
\begin{equation}\label{eq:secenh}
  \max_{s^n}\max_\gamma I(M\wedge Z^\gamma_{s^n})
\end{equation}
to be small. Both secrecy criteria assume that the eavesdropper knows the realization of the correlated randomness. This means that we have to assume the active and passive attacks to be uncoordinated in the sense that the eavesdropper does not inform the jammer about its knowledge of the correlated randomness.

We are not the first to study the capacity of the AVWC. A study of the Gaussian MIMO wiretap channel where the channel to the eavesdropper is arbitrarily varying has been done in \cite{HKY,HYa}. Earlier approaches to the discrete AVWC as defined in \eqref{eq:AVWCfamily} can be found in \cite{BBSa,MBL}, which studied the secrecy capacity achieved by correlated random coding and used \eqref{eq:secBBSa} as secrecy criterion. In both papers, closed-form secrecy capacity results could only be given after imposing additional conditions. 

The main result of this paper will be a complete characterization of the correlated randoom coding secrecy capacity under both criteria \eqref{eq:secBBSa} and \eqref{eq:secenh}. The capacity formula we find is multi-letter.\label{intro:multi-letter} It was found in \cite{BBSa} for special AVWCs where there is a ``best channel to the eavesdropper'' and reduces to a single-letter formula under certain degradedness conditions as required in \cite{MBL}. It is not clear whether a generally applicable single-letter formula exists at all. Still, the multi-letter formula allows for the approximate computation of the secrecy capacity up to a given complexity. However, this is not our main concern, so we do not provide any relation between complexity and approximation goodness. 

With the help of the multi-letter formula, it can also be shown that the correlated random coding secrecy capacity is continuous in the channel. Thus small errors in the description of the family \eqref{eq:AVWCfamily} do not have severe consequences on the capacity. If the capacity formula were not continuous, the channel would in general have to be estimated with infinite precision in order to meaningfully apply the capacity formula. The continuity of the correlated random coding secrecy capacity becomes even more remarkable as very simple examples with $\lvert\mathcal S\rvert=2$ have been given in \cite{BSP} which show that the uncorrelated coding secrecy capacity is a discontinuous function of the AVWC. 

For the achievability part of the capacity theorem, we follow Ahlswede's strategy of deriving correlated random coding achievability results for AVCs from uncorrelated coding capacity results for compound channels. (In contrast to an AVC, a compound channel does not change its state during the transmission of a codeword.) This technique is known as the ``robustification technique''. Sender and receiver of an AVC randomly permute an uncorrelated code for a certain compound channel induced by the AVC and thus obtain a correlated random code with negligibly larger average error.

When applying the robustification technique to AVWCs, one has to take the secrecy criterion into account. As seen in \cite{BBSa}, this requires a ``best channel to the eavesdropper'' if one assumes the channel to the eavesdropper to be compound as well. The central idea of our proof is to introduce the compound-arbitrarily varying wiretap channel (CAVWC). This channel is compound from sender to intended receiver and arbitrarily varying from sender to eavesdropper. We derive the uncorrelated coding secrecy capacity of this channel. After robustification, this also turns out to be the correlated random coding secrecy capacity of the AVWC.

We prove the achievability result for the CAVWC by random coding following Devetak \cite{De05}. This technique takes a resolvability approach to proving secrecy, cf.\ the discussion of resolvability and ``capacity-based" approaches by Bloch and Laneman \cite{BLa}. However, it does not follow an information spectrum approach like the techniques presented in \cite{BLa}. To our knowledge, those techniques have not yet been shown to be able to handle arbitrarily varying channels. As the number of AVWC channel states grows exponentially with blocklength, very tight probability estimates have to be obtained from random coding. Devetak's method \cite{De05}, originally in the language of quantum information theory, provides such estimates and was already applied in \cite{WWT} in a classical information theory setting.

In \cite{BS}, an a priori upper bound on the amount of correlated randomness required to achieve the correlated random coding secrecy capacity was found. Such a bound is necessary for the converse of the correlated random coding secrecy capacity theorem for the AVWC. The reason for this is that the use of correlated randomness prohibits a straightforward application of the data processing inequality. 

In a follow-up work \cite{AVWCII} to this paper, the AVWC correlated random coding secrecy capacity for the case that the eavesdropper has no knowledge of the correlated randomness as well as the AVWC uncorrelated coding secrecy capacity are studied.

\textit{Paper outline:} In Section II, we set the notation and give basic definitions. In Section III we define the AVWC and state the coding problem and the main result. Section IV discusses the main result of Section III. Section V introduces the CAVWC mentioned in the introduction, states the CAVWC coding problem and the corresponding secrecy capacity theorem. Section VI contains the proof of the achievability part of the coding theorem for the CAVWC. The achievability part of the correlated random coding theorem for the AVWC is derived from the achievability part of the coding theorem for the CAVWC in Section VII. Section VIII contains the converses. In Section IX, a short discussion concludes the paper. Several proofs are collected in the appendices.

\section{Notation and Basic Definitions}

Logarithms denoted by $\log$ are taken to the base $2$; correspondingly, we set $\exp(x)=2^x$. The cardinality of a finite set $\mathcal A$ is written $\lvert \mathcal A\rvert$. For a subset $\mathcal E$ of $\mathcal A$, we write $\mathcal E^c:=\mathcal A\setminus\mathcal E$. The \emph{indicator function} $\mathbbm 1_{\mathcal E}$ assumes the value $1$ for arguments contained in $\mathcal E$ and $0$ else. For $n$-tuples contained in $\mathcal A^n$, we write $x^n:=(x_1,\ldots,x_n)\in\mathcal A^n$.

The set of probability measures on the finite set $\mathcal A$ is denoted by $\mathcal P(\mathcal A)$. For $P\in\mathcal P(\mathcal A)$, we define the $n$-fold product measure $P^n\in\mathcal P(\mathcal A^n)$ by $P^n(x^n):=\prod_i P(x_i)$. We write stochastic matrices $\{W(b\vert a):a\in\mathcal A,b\in\mathcal B\}$ with input alphabet $\mathcal A$ and output alphabet $\mathcal B$ as mappings $W:\mathcal A\longrightarrow\mathcal P(\mathcal B)$. A nonnegative measure on $\mathcal A$ is a vector $(\mu(a))_{a\in\mathcal A}$ with $\mu(a)\geq0$ for all $a\in\mathcal A$. A probability measure is a nonnegative measure. The total variation distance of two nonnegative measures $\mu,\nu$ on $\mathcal A$ is defined by $\lVert\mu-\nu\rVert:=\sum_{a\in\mathcal A}\lvert \mu(a)-\nu(a)\rvert$.

If $\bar X,\bar Y$ are random variables, then we write the distribution of $\bar X$ as $P_{\bar X}$, the joint distribution of $\bar X$ and $\bar Y$ as $P_{\bar X\bar Y}$ and the conditional distribution of $\bar X$ given $\bar Y$ as $P_{\bar X\vert\bar Y}$.

For a sequence $x^n=(x_1,\ldots,x_n)\in\mathcal A^n$ and $a\in\mathcal A$, the number $N(a\vert x^n)$ indicates the number of coordinates $x_i$ of $x^n$ with $x_i=a$. The type of $x^n$ is the probability measure $q\in\mathcal P(\mathcal A)$ defined by $q(a):=N(a\vert x^n)/n$. The set of all possible types of sequences of length $n$ is denoted by $\mathcal P_0^n(\mathcal A)$. For $\delta>0$ and an $\mathcal A$-valued random variable $\bar X$, we define the typical set $\mathcal T_{\bar X,\delta}^n\subset\mathcal A^n$ as the set of those $x^n\in\mathcal A^n$ satisfying the two conditions
\begin{align*}
  \left\lvert \frac{1}{n}N(a\vert x^n)-P_{\bar X}(a)\right\rvert<\delta\qquad\text{for every }a\in\mathcal A,\\
  N(a\vert x^n)=0\text{ if }P_{\bar X}(a)=0.
\end{align*}
For $\delta>0$, an $\mathcal A\times\mathcal B$-valued random variable $(\bar X,\bar Y)$ with joint distribution $P_{\bar X\bar Y}$ and an element $x^n$ of $\mathcal A^n$, we define the conditionally typical set $\mathcal T_{\bar Y\vert\bar X,\delta}^n(x^n)$ as the set of those $y^n\in\mathcal B^n$ satisfying the two conditions
\begin{align*}
  \left\lvert\frac{1}{n}N(a,b\vert x^n,y^n)-P_{\bar Y\vert\bar X}(b\vert a)\frac{1}{n}N(a\vert x^n)\right\rvert<\delta\qquad\text{for all }a\in\mathcal A,b\in\mathcal B,\\
  N(a,b\vert x^n,y^n)=0\text{ if }P_{\bar Y\vert\bar X}(b\vert a)=0.
\end{align*}

\section{Arbitrarily Varying Wiretap Channels}

Let $\mathcal A,\mathcal B,\mathcal C,\mathcal S$ be finite sets. For every $s\in\mathcal S$, let a stochastic matrix $W_s:\mathcal A\rightarrow\mathcal{P}(\mathcal B)$ and another stochastic matrix $V_s:\mathcal A\rightarrow\mathcal{P}(\mathcal C)$ be given. For a number $n$ and $x^n\in\mathcal A^n,y^n\in\mathcal B^n,s^n\in\mathcal S^n$, define
\[
  W^n_{s^n}(y^n\vert x^n):=\prod_{i=1}^nW_{s_i}(y_i\vert x_i).
\]
We denote the family $\{W^n_{s^n}:s^n\in\mathcal S^n,n=1,2,\ldots\}$ by $\mathfrak{W}$. In analogy to $W_{s^n}^n(y^n\vert x^n)$, we define $V^n_{s^n}(z^n\vert x^n)$ for $z^n\in\mathcal C^n$ and denote the corresponding family $\{V_{s^n}^n:s^n\in\mathcal S^n,n=1,2,\ldots\}$ by $\mathfrak V$. We sometimes prefer to write $V^n(z^n\vert x^n,s^n)$ instead of $V_{s^n}^n(z^n\vert x^n)$. We call the pair $(\mathfrak{W,V})$ an \emph{Arbitrarily Varying Wiretap Channel (AVWC)}. $\mathcal S$ is called the \emph{state set} of $(\mathfrak{W,V})$.

\begin{rem}\label{rem:joint}
  One checks easily that the representation of an AVWC as a pair $(\mathfrak W,\mathfrak V)$ is possible without losing generality. In general, any state $s\in\mathcal S$ together with an input $a\in\mathcal A$ will lead to a joint output distribution $U_s(\cdot,\cdot\vert a)$. But the performance of any of the codes defined below is measured with respect to the marginal output distributions $W_s(\cdot\vert a)$ and $V_s(\cdot\vert a)$. Thus for the purpose of this paper, all AVWCs with the same marginals $\mathfrak W$ and $\mathfrak V$ are equivalent.
\end{rem}

An uncorrelated $(n,J_n)$-code $\mathcal{K}_n$ for the AVWC $(\mathfrak{W,V})$ consists of a stochastic encoder $E:\{1,\ldots,J_n\}\rightarrow\mathcal{P}(\mathcal A^n)$ and a collection of mutually disjoint sets $\{\mathcal D_j\subset\mathcal B^n:1\leq j\leq J_n\}$ whose union equals $\mathcal B^n$. We abbreviate $\mathcal{J}_n:=\{1,\ldots,J_n\}$. Together with an AVWC $(\mathfrak{W,V})$, any uncorrelated $(n,J_n)$-code $\mathcal{K}_n$ defines a canonical family 
\begin{equation}\label{eq:canonnopi}
	\mathcal F(\mathcal K_n,\mathfrak{W,V}):=\{M^n,X^n,Y_{s^n}^n,Z_{s^n}^n,\hat M_{s^n}^n:s^n\in\mathcal S^n\}
\end{equation}
of random variables, with $M^n$ and $\hat M_{s^n}^n$ assuming values in $\mathcal J_n$, the values of $X^n$ in $\mathcal A^n$, those of $Y_{s^n}^n$ in $\mathcal B^n$, those of $Z_{s^n}^n$ in $\mathcal C^n$, and such that for every $s^n\in\mathcal S^n$ the distribution of $(M^n,X^n,Y_{s^n}^n,Z_{s^n}^n,\hat M_{s^n}^n)$ equals
\[
	P_{M^nX^nY_{s^n}^nZ_{s^n}^n\hat M_{s^n}^n}(j,x^n,y^n,z^n,\hat\jmath)=\frac{1}{J_n}E(x^n\vert j)W_{s^n}^n(y^n\vert x^n)V_{s^n}^n(z^n\vert x^n)\mathbbm 1_{\mathcal D_{\hat\jmath}}(y^n).
\]
Recall that we incur no loss of generality by defining $Y_{s^n}^n$ and $Z_{s^n}^n$ to be independent conditional on $X^n$, as the joint distribution of $Y_{s^n}^n$ and $Z_{s^n}^n$ will never play any role (cf. Remark \ref{rem:joint}). The average error of $\mathcal{K}_n$ is given by
\begin{align*}
  e(\mathcal{K}_n)&:=\max_{s^n\in\mathcal S^n}\mathbb P[M^n\neq\hat M_{s^n}^n].
\end{align*}

\begin{defn}\label{def:uncAVWC}
  A non-negative number $R_S$ is an \emph{achievable uncorrelated coding secrecy rate for the AVWC $(\mathfrak{W,V})$} if there exists a sequence $(\mathcal{K}_n)_{n=1}^\infty$ of uncorrelated $(n,J_n)$-codes such that
\begin{align}
  \liminf_{n\rightarrow \infty}\frac{1}{n}\log J_n&\geq R_S,\\
  \lim_{n\rightarrow\infty} e(\mathcal{K}_n)&=0,\label{eq:aerr}\\
  \lim_{n\rightarrow\infty}\max_{s^n\in\mathcal S^n}I(M^n\wedge Z_{s^n}^n)&=0.\label{eq:sec}
\end{align}
The \emph{uncorrelated coding secrecy capacity of $(\mathfrak{W,V})$} is the supremum of all achievable secrecy rates $R_S$ and is denoted by $C_S(\mathfrak{W,V})$.
\end{defn}

Note the different roles the families $\mathfrak W$ and $\mathfrak V$ play. $\mathfrak W$ is an \emph{Arbitrarily Varying Channel (AVC)} from a sender with alphabet $\mathcal A$ to a receiver with alphabet $\mathcal B$. Messages are supposed to be sent over this AVC in such a way that only a small, asymptotically negligible average error is incurred. This is reflected in condition \eqref{eq:aerr}. This communication is subject to an additional secrecy condition. An eavesdropper obtains a noisy version of the sender's channel inputs via the AVC $\mathfrak V$. Condition \eqref{eq:sec} guarantees secrecy no matter what the channel state is.

For given $(n,J_n)$, we assume that the set of uncorrelated $(n,J_n)$-codes is indexed by the set $\Gamma_n$. That means that the set of all uncorrelated $(n,J_n)$-codes (with given channel input and output alphabets $\mathcal A$ and $\mathcal B$) has the form $\{\mathcal K_n(\gamma):\gamma\in\Gamma_n\}$. For the uncorrelated $(n,J_n)$-code $\mathcal K_n(\gamma)$, with $\gamma\in\Gamma_n$, we write for the canonical family of random variables 
\[
	\mathcal F(\mathcal K_n(\gamma),\mathfrak{W,V})=\{M^n,X^n(\gamma),Y_{s^n}^n(\gamma),Z_{s^n}^n(\gamma),\hat M_{s^n}^n(\gamma):s^n\in\mathcal S^n,\gamma\in\Gamma_n\}.
\] 
A correlated random $(n,J_n)$-code $\mathcal{K}_n^\ran$ for the AVWC $(\mathfrak{W,V})$ then is given by a finitely supported\footnote{``Finitely supported" means that the set $\supp(G_n):=\{\gamma\in\Gamma_n:P_{G_n}(\gamma)>0\}$ called the \textit{support of $G_n$} is finite.} random variable $G_n$ on $\Gamma_n$ independent of all canonical families of random variables $\mathcal F(\mathcal K_n(\gamma),\mathfrak{W,V})$. In other words, $G_n$ randomly chooses an uncorrelated $(n,J_n)$-code out of all possible ones and is independent of the message random variable, the randomness in the chosen stochastic encoder and the channel noise. The average error $e(\mathcal{K}_n^\ran)$ is defined as
\[
  e(\mathcal{K}_n^\ran):=\max_{s^n\in\mathcal S^n}\mathbb P[M^n\neq\hat M_{s^n}^n(G_n)]=\max_{s^n\in\mathcal S^n}\sum_{\gamma\in\Gamma_n}\mathbb P[M^n\neq\hat M_{s^n}^n(\gamma)]P_{G_n}(\gamma),
\]
where $\sum_{\gamma\in\Gamma_n}a(\gamma)P_{G_n}(\gamma)$ is short for the finite sum $\sum_{\gamma\in\supp(G_n)}a(\gamma)P_{G_n}(\gamma)$.

In the case of correlated random codes, we consider two secrecy criteria, leading to two different notions of achievable rate. 

\begin{defn}\label{def:corrmean}
  A non-negative number $R_S$ is called an \emph{achievable correlated random coding mean secrecy rate for the AVWC $(\mathfrak{W,V})$} if there exists a sequence $(\mathcal{K}_n^\ran)_{n=1}^\infty$ of correlated random $(n,J_n)$-codes such that
\begin{align}
  \liminf_{n\rightarrow\infty}\frac{1}{n}\log J_n&\geq R_S,\label{eq:corrrate}\\
  \lim_{n\rightarrow\infty}e(\mathcal{K}_n^\ran)&=0,\label{eq:corraerr}\\
  \lim_{n\rightarrow\infty}\max_{s^n\in\mathcal S^n}I(M^n\wedge Z_{s^n}^n(G_n)\vert G_n)&=0.\label{eq:corrsec}
\end{align}
The supremum of all achievable secrecy rates for correlated random codes is called the \emph{correlated random coding mean secrecy capacity of $(\mathfrak{W,V})$} and denoted by $C^\mean_{S,\ran}(\mathfrak{W,V})$.
\end{defn}

\begin{defn}\label{def:corrmax}
  A non-negative number $R_S$ is called an \emph{achievable correlated random coding maximal secrecy rate for the AVWC $(\mathfrak{W,V})$} if there exists a sequence $(\mathcal{K}_n^\ran)_{n=1}^\infty$ of correlated random $(n,J_n)$-codes such that \eqref{eq:corrrate} and \eqref{eq:corraerr} hold and
\begin{equation}\label{eq:enhsec}
	\lim_{n\rightarrow\infty}\max_{s^n\in\mathcal S^n}\max_{\gamma\in\supp(G_n)}I(M^n\wedge Z_{s^n}^n(\gamma))=0.
\end{equation}
The supremum of all achievable correlated random coding maximal secrecy rates is called the \emph{correlated random coding maximal secrecy capacity of $(\mathfrak{W,V})$} and denoted by $C^{\max}_{S,\ran}(\mathfrak{W,V})$.
\end{defn}

\begin{rem}\label{rem:AVWCKapVergl}
	It is immediately clear that $C^\mean_{S,\ran}(\mathfrak{W,V})\geq C^{\max}_{S,\ran}(\mathfrak{W,V})$.
\end{rem}

The secrecy capacities for correlated random codes are characterized by a multi-letter formula, extending the results of \cite{BBSa}. We set
\begin{equation}\label{eq:R_S}
  R_S^*(\mathfrak{W,V})\\:=\lim_{k\rightarrow\infty}\frac{1}{k}\sup_{\{\bar U,\bar X^k,\bar Y_q^k,\bar Z_{s^k}^k\}}\Bigl(\min_{q\in\mathcal P(\mathcal S)}I(\bar U\wedge\bar  Y_q^k)-\max_{s^k\in\mathcal S^k}I(\bar U\wedge\bar  Z_{s^k}^k)\Bigr)
\end{equation}
where the supremum is over the set of families of random variables
\begin{equation}\label{eq:optset}
	\{\bar U,\bar X^k,\bar Y_q^k,\bar Z_{s^k}^k:q\in\mathcal P(\mathcal S),s^k\in\mathcal S^k\}
\end{equation}
satisfying that $\bar U$ assumes values in some finite subset of the integers, the values of $\bar X^k$ lie in $\mathcal A^k$, those of $\bar Y_q^k$ in $\mathcal B^k$, those of $\bar Z_{s^k}^k$ in $\mathcal C^k$, and such that for every $q\in\mathcal P(\mathcal S)$ and $s^k\in\mathcal S^k$, 
\begin{equation}\label{eq:canondist}
	P_{\bar U\bar X^k\bar Y_q^k\bar Z_{s^k}^k}(u,x^k,y^k,z^k)=P_{\bar U}(u)P_{\bar X^k\vert\bar  U}(x^k\vert u)\left(\prod_{i=1}^k\left[\sum_{s\in\mathcal S}q(s)W_s(y_i\vert x_i)\right]\right)V_{s^k}^k(z^k\vert x^k).
\end{equation}
$P_{\bar U}$ and $P_{\bar X^k\vert\bar  U}$ may be arbitrary probability distributions and stochastic matrices, respectively.

\begin{thm}\label{thm:capacity}
For the AVWC $(\mathfrak{W,V})$, we have
\[
	C^\mean_{S,\ran}(\mathfrak{W,V})=C^{\max}_{S,\ran}(\mathfrak{W,V})=R_S^*(\mathfrak{W,V}).
\]
\end{thm}

\begin{rem}\label{rem:remCAVWC}
\begin{enumerate}
	\item It is shown exactly as in \cite{Bjelakovic2013}, using Fekete's lemma \cite{Fek}, that the limit on the right-hand side of \eqref{eq:R_S} indeed exists. In fact, the limit can be replaced by a supremum, as the terms $\frac{1}{k}\sup(\ldots)$ increase in $k$.\label{rem:remCAVWC-limsup}
	\item For given $k$, the cardinality of $\mathcal U$ can be restricted to $\lvert \mathcal A\rvert^k$. This can be proved almost exactly as in the proof of \cite[Theorem 17.11]{CK}. The supremum in \eqref{eq:R_S} then becomes a maximum. \label{rem:remCAVWC-card}
	\item If for $q\in\mathcal P(\mathcal S)$ we define $W_q(b\vert a):=\sum_sq(s)W_s(b\vert a)$, the conditional probability of $\bar Y_q^k$ given $\bar X^k$ in \eqref{eq:canondist} satisfies
\[
	P_{\bar Y_q^k\vert\bar X^k}(y^k\vert x^k)=\prod_{k=1}^kW_q(y_i\vert x_i)=:W_q^k(y^k\vert x^k).
\]
The family $\{W_q^n:q\in\mathcal P(\mathcal S),n=1,2,\ldots\}$ is a memoryless channel which does not change its state during the transmission of a codeword. Such channels will appear later under the name of \textit{compound channel}.
	\item The work \cite{AVWCII} following up on this paper makes use of the fact that 
	\begin{align}
		R_S^*(\mathfrak{W,V})&=\lim_{k\rightarrow\infty}\frac{1}{k}\sup_{\{\bar U,\bar X^k,\bar Y_{\tilde q},\bar Z_{s^k}^k\}}\Bigl(\min_{\tilde q\in\mathcal P(\mathcal S^k)}I(\bar U\wedge\bar  Y_{\tilde q}^k)-\max_{s^k\in\mathcal S^k}I(\bar U\wedge\bar  Z_{s^k}^k)\Bigr)\label{eq:ersteabw}\\
		&=\lim_{k\rightarrow\infty}\frac{1}{k}\sup_{\{\bar U,\bar X^k,\bar Y_{\tilde q_1},\bar Z_{\tilde q_2}^k\}}\Bigl(\min_{\tilde q_1\in\mathcal P(\mathcal S^k)}I(\bar U\wedge\bar  Y_{\tilde q_1}^k)-\max_{\tilde q_2\in\mathcal P(\mathcal S^k)}I(\bar U\wedge\bar  Z_{\tilde q_2}^k)\Bigr),\label{eq:zweiteabw}
	\end{align}
	where the family of random variables in \eqref{eq:zweiteabw} is defined analogously to the family \eqref{eq:optset} with the difference that the parameters $\tilde q_1,\tilde q_2$ range over all probability distributions on $\mathcal P(\mathcal S^k)$ (in particular, not just the product measures with constant marginals or the extremal Dirac distributions) and where for $\tilde q_1,\tilde q_2\in\mathcal P(\mathcal S^k)$
	\[
		P_{\bar Y_{\tilde q_1}^k\bar Z_{\tilde q_2}^k\vert \bar X}(y^k,z^k\vert x^k)=\left(\sum_{s^k}\tilde q_1(s^k)W_{s^k}^k(y^k\vert x^k)\right)\left(\sum_{s^k}\tilde q_2(s^k)V_{s^k}^k(z^k\vert x^k)\right).
	\]
	The family of random variables in \eqref{eq:ersteabw} over which the supremum is taken is obtained by restricting the parameters $\tilde q_2$ in the family of random variables in \eqref{eq:zweiteabw} to the extremal Dirac measures, which means nothing else than to take $P_{\bar Z_{s^k}^k\vert\bar X}$ as in \eqref{eq:optset}. Similarly, by restricting the $\tilde q_1$ to be product measures on $\mathcal S^k$ with constant marginals, one can regard \eqref{eq:optset} itself as a restriction of the family in \eqref{eq:ersteabw}.
	
	To prove the equalities \eqref{eq:ersteabw} and \eqref{eq:zweiteabw}, first note that due to the convexity of mutual information in the channel nothing changes if $\max_{s^k\in\mathcal S^k}I(\bar U\wedge\bar  Z_{s^k})$ on the right-hand side of \eqref{eq:ersteabw} is replaced by $\max_{\tilde q_2\in\mathcal P(\mathcal S^k)}I(\bar U\wedge\bar  Z_{\tilde q_2}^k)$. This proves equality in \eqref{eq:zweiteabw}. It is also obvious that the right-hand side of \eqref{eq:ersteabw} is a lower bound on $R_S^*(\mathfrak{W,V})$. That equality holds can be seen by inspection of the proof of the converse in Section \ref{sect:conv} below. The main reason is the fact that the average decoding error for AVC and AVWC is affine in the channel, as proved in \cite[Lemma 12.3]{CK}. More details on this can be found in Remark \ref{rem:zusatz-conv} after the proof of the converse.
	
	The enlargement of the state space as in \eqref{eq:ersteabw} and \eqref{eq:zweiteabw} can be interpreted as allowing randomized jamming strategies. This does not affect the AVWC performance because the performance measures are robust against this randomization (i.~e.\ the average error is affine in the channel, mutual information between the message and the eavesdropper's output is even convex in the channel). \label{rem:remCAVWC-zusatz}
	\item Comparison of the right-hand side of \eqref{eq:R_S} with the capacity expressions derived in \cite{BLa} suggests that the terms $\min_{q\in\mathcal P(\mathcal S)}I(\bar U\wedge\bar Y_q^k)$ are related to an inf-information rate for the AVC $\mathfrak W$ and $\max_{s^k\in\mathcal S^k}I(\bar U\wedge\bar Z_{s^k}^k)$ to a sup-information rate for the AVC $\mathfrak V$, see also \cite{Han}. However, as AVCs have not yet been treated in the framework of the theory of information spectrum, this remains speculation for the time being.
\end{enumerate}
\end{rem}

\section{Discussion of Theorem \ref{thm:capacity}}

\subsection{Multi-letter vs.\ single-letter}\label{subsect:multi-letter}

The bound from Remark \ref{rem:remCAVWC}-\ref{rem:remCAVWC-card}) on the size of $\mathcal U$ for fixed $k$ does not give a general upper bound on the cardinality of the auxiliary alphabet $\mathcal U$. It could still be helpful in calculations of $R_S^*(\mathfrak{W,V})$ if one knows from other arguments that there exists a $k_0$ such that, for $k\geq k_0$,
\[
  \frac{1}{k}\sup_{\{\bar U,\bar X^k,\bar Y_q^k,\bar Z_{s^k}^k\}}\Bigl(I(\bar U\wedge \bar Y_q^k)-\max_{s^k\in\mathcal S^k}I(\bar U\wedge\bar  Z_{s^k}^k)\Bigr)
\]
is sufficiently close to $R_S^*(\mathfrak{W,V})$. From Remark \ref{rem:remCAVWC}-\ref{rem:remCAVWC-limsup}) it follows that this approach would give a lower bound on the secrecy capacity. Note that it is not at all clear whether a single-letter characterization of $R_S^*(\mathfrak{W,V})$ is available. In the case of the unavailability of a single-letter capacity expression, only approximate calculations of capacity are possible. 

That the above multi-letter characterization can lead to further insights into the nature of AVWCs can be seen in Subsection \ref{subsect:continuity}, where the continuity of $R_S^*(\mathfrak{W,V})$ in $(\mathfrak{W,V})$ is shown. To show this a priori, i.~e.\ without having the multi-letter expression for capacity, seems to be very hard. With the formula at hand, however, it can be done. For the uncorrelated coding secrecy capacity, a similar study of continuity is performed in \cite{AVWCII}, also on the basis of the multi-letter formula.

A single-letter formula for $C_{S,\ran}^\mean(\mathfrak{W,V})$ has been given in \cite{MBL} for AVWCs which satisfy certain conditions. We now present these conditions and show that if they are satisfied, the formula found in \cite{MBL} coincides with $R_S^*(\mathfrak{W,V})$, which then becomes single-letter.

The first condition of \cite{MBL} is that $(\mathfrak{W,V})$ be \textit{strongly degraded with independent states}. This means 
\begin{itemize}
	\item that $\mathcal S=\mathcal S_1\times\mathcal S_2$ and that the families $\{W_{(s_1,s_2)}:(s_1,s_2)\in\mathcal S_1\times\mathcal S_2\}$ and $\{V_{(s_1,s_2)}:(s_1,s_2)\in\mathcal S_1\times\mathcal S_2\}$ of stochastic matrices determining $\mathfrak{W}$ and $\mathfrak{V}$ satisfy $W_{(s_1,s_2)}=W_{s_1}$ and $V_{(s_1,s_2)}=V_{s_2}$ for all $(s_1,s_2)$; and
	\item that for every $q_1\in\mathcal P(\mathcal S_1)$ and $q_2\in\mathcal P(\mathcal S_2)$, the matrix $V_{q_2}$ should be a degraded version of $W_{q_1}$, where 
	\[
		W_{q_1}(y\vert x)=\sum_{s_1\in\mathcal S_1}W_{s_1}(y\vert x)q_1(s_1),\quad V_{q_2}(z\vert x)=\sum_{s_2\in\mathcal S_2}V_{s_2}(z\vert x)q_2(s_2),
	\]
	and $V_{q_2}$ is a degraded version of $W_{q_1}$ if there exists a stochastic matrix $T_{q_1q_2}:\mathcal B\rightarrow\mathcal C$ such that
	\begin{equation}\label{eq:degradedness}
		V_{q_2}(z\vert x)=\sum_yT_{q_1q_2}(z\vert y)W_{q_1}(y\vert x).
	\end{equation}
	(Observe: It is sufficient to require \eqref{eq:degradedness} to hold only for $s_2\in\mathcal S_2$ and $q_1\in\mathcal P(\mathcal S_1)$. The validity of \eqref{eq:degradedness} for all $q_1\in\mathcal P(\mathcal S_1)$ and $q_2\in\mathcal P(\mathcal S_2)$ then follows upon setting $T_{q_1q_2}(z\vert y):=\sum_{s_2}q_2(s_2)T_{q_1s_2}(z\vert y)$ for all $y\in\mathcal B,z\in\mathcal C$. Thus the function $(q_1,q_2)\mapsto T_{q_1q_2}$ can without loss of generality be assumed to be linear in $q_2$. This is not possible for $q_1$, as can be seen from analyzing Example 3 in \cite{MBL}.)
\end{itemize}
The second condition of \cite{MBL} is essentially the \textit{best channel to the eavesdropper} condition from \cite{BBSa}, so we will henceforth call it this way. It requires that there exists an $s_*\in\mathcal S_2$ such that for all $s_2\in\mathcal S_2$, the channel $V_{s_2}$ is a degraded version of $V_{s_*}$, with degradedness here defined analogously to \eqref{eq:degradedness}. (The general definition of ``best channel to the eavesdropper" in \cite{BBSa,MBL} does not require independent states.)

\begin{cor}\label{cor:MolavianJazi}
	If the AVWC $(\mathfrak{W,V})$ is strongly degraded with independent states and has a best channel to the eavesdropper, then 
	\begin{equation}\label{eq:MolavianJazi}
		R_S^*(\mathfrak{W,V})=\max_{\{\bar X,\bar Y_{q_1},\bar Z_{s_2}\}}\biggl(\min_{q_1\in\mathcal P(\mathcal S_1)}I(\bar X\wedge\bar  Y_{q_1})-\max_{s_2\in\mathcal S_2}I(\bar X\wedge\bar  Z_{s_2})\biggr)
	\end{equation}
	where the maximum over $\{\bar X,\bar Y_{q_1},\bar Z_{s_2}\}$ is over families of random values satisfying
	\[
		P_{\bar X\bar Y_{q_1}\bar Z_{s_2}}(x,y,z)=P_{\bar X}(x)W_{q_1}(y\vert x)V_{s_2}(z\vert x)
	\]
	and where $\bar X$ is an arbitrary $\mathcal A$-valued random variable.
\end{cor}

\begin{IEEEproof} 
See Appendix \ref{app:MolavianJazi}.
\end{IEEEproof}

\subsection{The amount of correlated randomness}\label{subsect:amCR}

Next we ask how many values the correlated randomness variable should attain with positive probability in order for $C_{S,\ran}^\mean(\mathfrak{W,V})$ and $C_{S,\ran}^{\max}(\mathfrak{W,V})$ to be achievable. This can be answered in an a priori fashion, so it can be applied in the converse of Theorem \ref{thm:capacity}.

Note that the definitions allow every kind of correlated randomness as long as it is finitely supported. In the achievability proof of Theorem \ref{thm:capacity}, we shall see that the uniform distribution on a set of cardinality $n!$ is sufficient, where $n$ is the blocklength of the code. The size of this set can still be reduced considerably. For AVCs, the first such reduction was presented by Ahlswede in \cite{A1}, where he showed that $\lvert\supp(G_n)\rvert\leq n^{1+\varepsilon}$ is sufficient. 

A stronger result has been found recently \cite{BS}. Its essence is that every secrecy rate $R_S<C^{\max}_{S,\ran}(\mathfrak{W,V})$ is achievable with no more than a finite amount of correlated randomness, given arbitrary upper bounds on the average error and the mutual information between message random variable and eavesdropper output. 

\begin{lem}[\cite{BS}]\label{lem:CRreduction}
	Let $R_S<C_{S,\ran}^\mean(\mathfrak{W,V})$ and $\lambda,\delta>0$. Then for every $\varepsilon>0$ there exists a positive integer $L=L(R_S,\varepsilon,\lambda,\delta)$ such that for sufficiently large $n$ there exists a correlated random $(n,J_n)$-code $\mathcal K_n^\ran$ satisfying
	\begin{align}
		\frac{1}{n}\log J_n&\geq R_S-\varepsilon,\label{eq:ratefiniteL}\\
		e(\mathcal K_n^\ran)&\leq\lambda,\\
		\max_{s^n\in\mathcal S^n}I(M^n\wedge Z_{s^n}^n(G_n)\vert G_n)&\leq\delta,\\
		\lvert\supp(G_n)\rvert&\leq L.\label{eq:finiteLfiniteL}
	\end{align}
\end{lem}
An analogous statement holds for $\max_{s^n\in\mathcal S^n}I(M^n\wedge Z_{s^n}^n(G_n)\vert G_n)$ replaced by $\max_{\gamma\in\supp(G_n)}\max_{s^n\in\mathcal S^n}I(M^n\wedge Z_{s^n}^n(\gamma))$. 

\subsection{Model robustness and continuity}\label{subsect:continuity}

Here we study the continuity of the correlated random coding secrecy capacity function in the channel. Continuity is an important property of a capacity function, a fact which is sometimes overlooked because single-letter formulas usually are obviously continuous. The question becomes non-trivial in the case of a multi-letter capacity formula like $R_S^*(\mathfrak{W,V})$.

Suppose the capacity function were not continuous and assume that one estimates a channel which is close to a point of discontinuity. Then this channel has to be estimated to a precision which might be higher than achievable in the estimation process, or even higher than a computer can handle with reasonable effort. Otherwise, the capacity expression obtained from the formula is next to useless for this particular channel, as all of its values in the neighbourhood of the estimated channel could be the correct one, and this range of possible values could take on arbitrary form. From this point of view, the lack of continuity of a capacity function is more dramatic than a lacking single-letter expression, because a multi-letter formula still allows an approximate calculation, whereas approximation is not possible if the capacity function is discontinuous.

We shall show that the capacity functions $C_{S,\ran}^\mean(\mathfrak{W,V})$ and $C_{S,\ran}^{\max}(\mathfrak{W,V})$ are continuous. The argumentation relies on the fact that we have an explicit formula for these, as $C_{S,\ran}^\mean(\mathfrak{W,V})=C_{S,\ran}^{\max}(\mathfrak{W,V})=R_S^*(\mathfrak{W,V})$. It is thus an example of the usefulness of a multi-letter formula.

 Of course, the set of AVWCs with given in- and output alphabets has to be equipped with a metric in order to be able to talk about the continuity of capacity in the channel. Let $(\mathfrak W,\mathfrak V)$ and $(\mathfrak{\tilde W},\mathfrak{\tilde V})$ be two AVWCs with input alphabet $\mathcal A$ and output alphabets $\mathcal B,\mathcal C$ for the legitimate receiver and the eavesdropper, respectively. Denote the finite state space of $(\mathfrak W,\mathfrak V)$ by $\mathcal S$ and the finite state space of $(\mathfrak{\tilde W},\mathfrak{\tilde V})$ by $\mathcal{\tilde S}$. We measure the distance of $(\mathfrak{W,V})$ and $(\mathfrak{\tilde W,\tilde V})$ by what is called the \emph{Hausdorff distance} of two sets.

For two stochastic matrices $W,\tilde W:\mathcal A\rightarrow\mathcal B$, we define
\[
  \lVert W-\tilde W\rVert_o:=\max_{a\in\mathcal A}\lVert W(\,\cdot\,\vert a)-\tilde W(\,\cdot\,\vert a)\rVert.
\]
We define four asymmetric distances
\begin{align*}
  d_{B,1}(\mathfrak W,\mathfrak{\tilde W})&:=\max_{\tilde s\in\mathcal{\tilde S}}\min_{s\in\mathcal S}\lVert W_s-\tilde W_{\tilde s}\rVert_o,\\
  d_{B,2}(\mathfrak W,\mathfrak{\tilde W})&:=\max_{s\in\mathcal S}\min_{\tilde s\in\mathcal{\tilde S}}\lVert W_s-\tilde W_{\tilde s}\rVert_o,
\end{align*}
and analogously define $d_{E,1}(\mathfrak V,\mathfrak{\tilde V}),d_{E,2}(\mathfrak V,\mathfrak{\tilde V})$ by replacing $W_s,\tilde W_{\tilde s}$ in the above definitions by $V_s,\tilde V_{\tilde s}$. Then the Hausdorff distance between $(\mathfrak{W,V}$ and $(\mathfrak{\tilde W,\tilde V})$ is defined by
\[
  d((\mathfrak W,\mathfrak V),(\mathfrak{\tilde W},\mathfrak{\tilde V})):=\max\bigl\{d_{B,1}(\mathfrak W,\mathfrak{\tilde W}),d_{E,1}(\mathfrak V,\mathfrak{\tilde V}),d_{B,2}(\mathfrak W,\mathfrak{\tilde W}),d_{E,2}(\mathfrak V,\mathfrak{\tilde V})\bigr\}.
\]
One checks easily that this is an actual metric on the set of finite-state AVWCs with the corresponding alphabets $\mathcal{A,B,C}$.

Building on Theorem \ref{thm:capacity}, we now state the central result concerning the continuity of the correlated random capacities.
 
\begin{thm}\label{lem:continuity}
  $R_S^*(\mathfrak W,\mathfrak V)$ is continuous in $(\mathfrak W,\mathfrak V)$ with respect to the metric $d$. Thus, $C_{S,\ran}(\mathfrak W,\mathfrak V)$ and $\hat C_{S,\ran}(\mathfrak W,\mathfrak V)$ are continuous functions of the channel.
\end{thm}

The proof of this theorem only requires minor changes compared to that of \cite[Theorem 2]{BSP} where the continuity the capacity of the corresponding compound wiretap channel is shown.

In contrast to the correlated random coding secrecy capacity, the uncorrelated coding secrecy capacity of AVWCs is known to be discontinuous. This was shown in \cite{BSP} with a very simple example on small alphabets and a state set of no more than two elements. Hence the continuity of the correlated random coding secrecy capacity becomes even more remarkable, especially as the previous subsection \ref{subsect:amCR} has shown that only very little correlated randomness is required to cause such a qualitative change of capacity functions. The exact characterization of the discontinuity points of the uncorrelated coding secrecy capacity $C_S(\mathfrak W,\mathfrak V)$ is more intricate. It is discussed in depth in \cite{AVWCII}.

\section{The Compound-Arbitrarily Varying Wiretap Channel}

To establish Theorem \ref{thm:capacity}, we use Ahlswede's robustification technique \cite{A3}. It was developed to turn deterministic codes for compound channels into correlated random codes for AVCs. It has already been applied in \cite{BBSa} to compound and arbitrarily varying wiretap channels. The difference of this paper's approach is that the channel from sender to eavesdropper will always be arbitrarily varying. Therefore it is no longer necessary to assume the existence of a best channel to the eavesdropper.

We now formalize the idea of having a compound channel from $\mathcal A$ to $\mathcal B$ and an arbitrarily varying channel from $\mathcal A$ to $\mathcal C$. Let $\mathcal R$ be any set. For every $r\in\mathcal R$, let $W_r:\mathcal X\longrightarrow\mathcal Y$ be a stochastic matrix. Set $W^n_r(y^n\vert x^n)=\prod_{i=1}^nW_r(y_i\vert x_i)$. Note that here, in contrast to the AVC, the channel state remains constant over time. This defines a \emph{compound channel} $\mathfrak{\overline W}:=\{W^n_r:r\in\mathcal R, n=1,2,\ldots\}$. Together with the AVC $\mathfrak V$ from the previous section, we obtain the \emph{compound-arbitrarily varying wiretap channel} (CAVWC) $(\mathfrak{\overline W,V})$.

We apply uncorrelated $(n,J_n)$-codes for message transmission over $(\mathfrak{\overline W},\mathfrak V)$. Together with $(\overline{\mathfrak{W}},\mathfrak{V})$, every $(n,J_n)$-code defines a canonical family of random variables
\begin{equation}\label{eq:canonCAVWC}
	\mathcal F(\mathcal K_n,\mathfrak{\overline W,V}):=\{(M^n,X^n,Y_r^n,Z_{s^n}^n,\hat M_r^n):r\in\mathcal R,s^n\in\mathcal S^n\},
\end{equation}
where $M^n$ and $\hat M_r^n$ assume values in $\mathcal J_n$, the values of $X^n$ lie in $\mathcal A^n$, those of $Y_r^n$ in $\mathcal B^n$ and those of $Z_{s^n}^n$ in $\mathcal C^n$ and where for any $r\in\mathcal R$ and $s^n\in\mathcal S^n$
\[
	P_{M^nX^nY_r^nZ_{s^n}^n\hat M_r^n}(j,x^n,y^n,z^n,\hat\jmath)=\frac{1}{J_n}E(x^n\vert j)W_r^n(y^n\vert x^n)V_{s^n}^n(z^n\vert x^n)\mathbbm 1_{\mathcal D_{\hat\jmath}}(y^n).
\]
For the uncorrelated $(n,J_n)$-code $\mathcal K_n$, the average error is defined as 
\[
  \bar e(\mathcal K_n):=\max_{r\in\mathcal R}\mathbb P[M^n\neq\hat M^n_r].
\]

\begin{defn}\label{def:ordinarysecrecyCAVWC}
  A nonnegative number $R_S$ is called an \emph{achievable secrecy rate for the CAVWC $(\mathfrak{\overline W,V})$} if there exists a sequence $(\mathcal{K}_n)_{n=1}^\infty$ of uncorrelated $(n,J_n)$-codes such that
\begin{align}
  \liminf_{n\rightarrow\infty}\frac{1}{n}\log J_n&\geq R_S,\notag\\
  \lim_{n\rightarrow\infty}\bar e(\mathcal{K}_n)&=0,\notag\\
  \lim_{n\rightarrow\infty}\max_{s^n\in\mathcal S^n}I(M^n\wedge Z_{s^n}^n)&=0.\label{eq:CAVWCsec}
\end{align}
The supremum of all achievable secrecy rates is called the \emph{secrecy capacity of $(\mathfrak{\overline W,V})$} and denoted by $C_{S}(\mathfrak{\overline W,V})$.
\end{defn}

We are actually interested in a stronger, permutation invariant form of secrecy. This is because we mainly consider CAVWCs as an auxiliary channel model. We would like to exploit the achievability part of a coding theorem for CAVWCs to find rates that are achievable for the AVWC by correlated random codes. This can be done using Ahlswede's robustification technique, which requires an exponential decrease of the average error and ``permutation invariance" of secrecy to be defined below.

For a permutation $\pi$ contained in the symmetric group $\Pi_n$ of permutations of $\{1,\ldots,n\}$, denote by $E^\pi$ the stochastic encoder obtained from a stochastic encoder $E$ via
\begin{equation}\label{eq:permut-enc}
  E^\pi(x^n\vert j):=E(\pi^{-1}(x^n)\vert j).
\end{equation}
Here, $\pi(x^n)=(x_{\pi(1)},\ldots,x_{\pi(n)})$ for any $x^n\in\mathcal A^n$. The corresponding decoding sets are $\mathcal D_j^\pi:=\{\pi(y^n):y^n\in\mathcal D_j\}$. This family of codes together with $(\overline{\mathfrak W},\mathfrak V)$ induces a canonical \textit{permutation-invariant} family of random variables
\begin{equation}\label{eq:canonpi}
	\mathcal F(\mathcal K_n,\mathfrak{\overline W,V},\Pi_n):=\{(M^n,X^n(\pi),Y_r^n(\pi),Z_{s^n}^n(\pi),\hat M_r^n(\pi)):r\in\mathcal R,s^n\in\mathcal S^n,\pi\in\Pi_n\},
\end{equation}
where $M^n$ and $\hat M_r^n(\pi)$ assume values in $\mathcal J_n$, the values of $X^n(\pi)$ lie in $\mathcal A^n$, those of $Y_r^n(\pi)$ in $\mathcal B^n$ and those of $Z_{s^n}^n(\pi)$ in $\mathcal C^n$ and where for any $r\in\mathcal R$ and $s^n\in\mathcal S^n$ and $\pi\in\Pi_n$
\[
	P_{M^nX^n(\pi)Y_r^n(\pi)Z_{s^n}^n(\pi)\hat M_r^n(\pi)}(j,x^n,y^n,z^n,\hat\jmath)=\frac{1}{J_n}E^\pi(x^n\vert j)W_r^n(y^n\vert x^n)V_{s^n}^n(z^n\vert x^n)\mathbbm 1_{\mathcal D_{\hat\jmath}^\pi}(y^n).
\]
For every permutation, we have $\mathbb P[M^n\neq\hat M^n(\pi)]=\mathbb P[M^n\neq\hat M^n(\id)]$, where $\id$ denoted the identity permutation. Thus also in the permutation-invariant setting, we can still just write $\bar e(\mathcal K_n)$ for the average error of $\mathcal K_n$.

\begin{defn}\label{def:pinv}
  A nonnegative number $R_S$ is called an \emph{achievable permutation invariant secrecy rate for the CAVWC $(\mathfrak{\overline W,V})$} if there exists a sequence $(\mathcal{K}_n)_{n=1}^\infty$ of uncorrelated $(n,J_n)$-codes and a $\beta>0$ such that
\begin{align}
  \liminf_{n\rightarrow\infty}\frac{1}{n}\log J_n&\geq R_S,\label{eq:enhrate}\\
  \limsup_{n\rightarrow\infty}-\frac{1}{n}\log\bar e(\mathcal{K}_n)&\geq\beta,\label{eq:enhbed1}\\
  \lim_{n\rightarrow\infty}\max_{s^n\in S^n}\max_{\pi\in\Pi_n}I(M^n\wedge Z_{s^n}^n(\pi))&=0.\label{eq:CAVWCsecenh}
\end{align}
The supremum of all achievable permutation invariant secrecy rates is called the \emph{permutation invariant secrecy capacity of $(\mathfrak{\overline W,V})$} and denoted by $C^\pinv_{S}(\mathfrak{\overline W,V})$.
\end{defn}

\begin{thm}\label{thm:lowerboundCAVWC}
  The permutation invariant secrecy capacity $C^\pinv_S(\mathfrak{\overline W,V})$ and the secrecy capacity $C_S(\mathfrak{\overline W,V})$ of the CAVWC $(\mathfrak{\overline W,V})$ both equal
\[
  R_S^*(\mathfrak{\overline W,V})\\:=\lim_{k\rightarrow\infty}\frac{1}{k}\sup_{\{\bar U,\bar X^k,\bar Y_r^k,\bar Z_{s^k}^k\}}\Bigl(\min_{r\in\mathcal R}I(\bar U\wedge\bar Y_r^k)-\max_{s^k\in\mathcal S^k}I(\bar U\wedge\bar Z_{s^k}^k)\Bigr),
\]
where the supremum is over the set of families of random variables
\[
	\{\bar U,\bar X^k,\bar Y_r^k,\bar Z_{s^k}^k:r\in\mathcal R,s^k\in\mathcal S^k\}
\]
satisfying that $\bar U$ assumes values in a finite subset of the integers, the values of $\bar X^k$ lie in $\mathcal A^k$, those of $\bar Y_r^k$ in $\mathcal B^k$, those of $\bar Z_{s^k}^k$ in $\mathcal C^k$, and such that for every $r\in\mathcal R$ and $s^k\in\mathcal S^k$, 
\[
	P_{\bar U\bar X^k\bar Y_r^k\bar Z_{s^k}^k}(u,x^k,y^k,z^k)=P_{\bar U}(u)P_{\bar X^k\vert\bar  U}(x^k\vert u)W_{r}^k(y^k\vert x^k)V_{s^k}^k(z^k\vert x^k).
\]
$P_{\bar U}$ and $P_{\bar X\vert\bar  U}$ may be arbitrary probability distributions and stochastic matrices, respectively.
\end{thm}

Remarks \ref{rem:remCAVWC}-1), \ref{rem:remCAVWC}-2), \ref{rem:remCAVWC}-4) and \ref{rem:remCAVWC}-5) apply here as well.

\section{Achievability part of the proof of Theorem \ref{thm:lowerboundCAVWC}}\label{sect:CAVWCach}

\subsection{Reduction}

\begin{itemize}
	\item As $C_S^\pinv(\overline{\mathfrak{W}},\mathfrak{V})\leq C_S(\overline{\mathfrak{W}},\mathfrak{V})$, it is sufficient to show that $R_S^*(\mathfrak{\overline W,V})$ is an achievable permutation invariant secrecy rate for $(\mathfrak{W,V})$. 

\item Call $R_S\geq 0$ an \textit{achievable secrecy rate with exponentially decreasing error for the CAVWC $(\mathfrak{\overline W,V})$} if there exists a sequence $(\mathcal{K}_n)_{n=1}^\infty$ of uncorrelated $(n,J_n)$-codes and a $\beta>0$ such that
\begin{align}
  \liminf_{n\rightarrow\infty}\frac{1}{n}\log J_n&\geq R_S,\label{eq:permrate}\\
  \limsup_{n\rightarrow\infty}-\frac{1}{n}\log\bar e(\mathcal{K}_n)&\geq\beta,\label{eq:permerr}\\
  \lim_{n\rightarrow\infty}\max_{s^n\in S^n}I(M^n\wedge Z_{s^n}^n)&=0,\label{eq:permsec}
\end{align}
where $M^n$ and the $Z_{s^n}^n$ are the corresponding elements of $\mathcal F(\mathcal K_n,\overline{\mathfrak{W}},\mathfrak{V})$. It is sufficient to prove that $R_S^*(\mathfrak{\overline W,V})$ is an achievable secrecy rate with exponentially decreasing error for $(\mathfrak{\overline W,V})$. This is due to the following lemma.

\begin{lem}\label{lem:permutations}
	Let $\mathcal K_n$ be an uncorrelated $(n,J_n)$-code with stochastic encoder $E$. Let $M^n$ be the canonical message random variable and $\{Z_{s^n}^n(\pi):s^n\in\mathcal S^n,\pi\in\Pi_n\}$ the family of canonical eavesdropper output random variables from $\mathcal F(\mathcal K_n,\mathfrak{\overline W,V},\Pi_n)$. Let $\id$ be the identity permutation mapping each element of $\{1,\ldots,n\}$ to itself. If there exists an $\varepsilon>0$ such that
	\begin{equation}\label{eq:gegeben}
		\max_{s^n}I(M^n\wedge Z_{s^n}^n(\id))\leq\varepsilon,
	\end{equation}
	then
	\begin{equation}\label{eq:gewollt}
		\max_{\pi\in\Pi_n}\max_{s^n}I(M^n\wedge Z_{s^n}^n(\pi))\leq\varepsilon.
	\end{equation}
\end{lem}

Lemma \ref{lem:permutations} is proved in Appendix \ref{app:permutations} and bases on the fact that $P_{M^n\pi(Z_{s^n}^n(\id))}=P_{M^nZ_{\pi(s^n)}^n(\pi)}$.

	\item $R_S^*(\mathfrak{\overline W,V})$ is an achievable secrecy rate with exponentially decreasing error if, for every CAVWC $(\mathfrak{\overline W,V})$, the rate 
	\begin{equation}\label{eq:ersterate}
  R_S^\dagger(\mathfrak{\overline W,V}):=\max_{\{\bar X,\bar Y_r,\bar Z_q\}}\Bigl(\min_{r\in\mathcal R}I(\bar X\wedge\bar  Y_r)-\max_{q\in\mathcal P(\mathcal S)}I(\bar X\wedge\bar  Z_q)\Bigr),
\end{equation}
is an achievable secrecy rate with exponentially decreasing error for $(\mathfrak{\overline W,V})$, where the maximum is over families of random variables $\{\bar X,\bar Y_r,\bar Z_q:r\in\mathcal R,q\in\mathcal P(\mathcal S)\}$, with $\bar X$ an arbitrary random variable assuming values in $\mathcal A$, the values of $\bar Y_r$ in $\mathcal B$, those of $\bar Z_q$ in $\mathcal C$, and 
\[
	P_{\bar X\bar Y_r\bar Z_q}(x,y,z)=P_{\bar X}(x)W_r(y\vert x)\left(\sum_{s\in\mathcal S}q(s)V_s(z\vert x)\right).
\]
This is proved using a standard channel prefixing argument, see Appendix \ref{app:prefixrand}.
\end{itemize}

\subsection{$R_S^\dagger(\mathfrak{\overline W,V})$ is an achievable secrecy rate with exponentially decreasing error}\label{subsect:exp-ach-dagger}

The proof that $R_S^\dagger(\mathfrak{\overline W,V})$ is an achievable secrecy rate with exponentially decreasing error for $(\mathfrak{\overline W,V})$ follows a random coding strategy. The random codewords are chosen as follows. Fix a blocklength $n$ and a family $\{\bar X,\bar Y_r,\bar Z_q:r\in\mathcal R,q\in\mathcal P(\mathcal S)\}$ as in the definition of $R_S^\dagger(\mathfrak{\overline W,V})$. For arbitrary $\tau>0$, set\footnote{Recall that we use the convention $\exp(x)=2^x$.}
\begin{align}
  J_n&:=\left\lfloor\exp\Bigl\{n\bigl(\min_{r\in\mathcal R}I(\bar X\wedge\bar Y_r)-\max_{q\in\mathcal P(S)}I(\bar X\wedge\bar Z_q)-\tau)\Bigr\}\right\rfloor,\label{eq:J_n-Def}\\
  L_n&:=\left\lfloor\exp\Bigl\{n\max_{q\in\mathcal P(S)}I(\bar X\wedge\bar Y_r)+\frac{\tau}{4}\Bigr\}\right\rfloor.\notag
\end{align}
and define $\mathcal J_n=\{1,\ldots,J_n\}$ and $\mathcal{L}_n:=\{1,\ldots,L_n\}$. Further, for some $\delta>0$ to be chosen later, we define a family $\mathcal X:=\{X_{jl}:j\in\mathcal J_n,l\in\mathcal L_n\}$ of random codewords in $\mathcal X^n$ with distribution\label{eq:unnumbered}
\[
  \mathbb P[X_{jl}=x^n]:=P'(x^n):=\frac{P_{\bar X}^n(x^n)}{P_{\bar X}^n(\mathcal T_{\bar X,\delta}^n)}\mathbbm 1_{\mathcal T_{\bar X,\delta}^n}(x^n).
\]
Via $\mathcal X$, we obtain a randomly selected stochastic encoder
\begin{equation}\label{eq:stochenc}
  E^{\mathcal X}(x^n\vert j):=\frac{1}{L_n}\sum_{l=1}^{L_n}\mathbbm 1_{\{X_{jl}\}}(x^n).
\end{equation}

\subsubsection{Reliability}

With high probability, a realization of $E^{\mathcal X}$ determines an uncorrelated $(n,J_n)$-code $\mathcal K_n^\ran$ for the compound channel $\overline{\mathfrak W}$ with exponentially small average error.

\begin{lem}\label{lem:transmission}
  For sufficiently small $\delta>0$ there exists a $\tau_6>0$ such that, if $n$ is sufficiently large, there exist decoding sets $\{\mathcal D_{j}^{\mathcal X}:j\in\mathcal J_n\}$ depending on $\mathcal X$ such that the event
\[
  \iota_3:=\left\{\sup_{r\in\mathcal R}\frac{1}{J_n}\sum_{j}\sum_{x^n}E^{\mathcal X}(x^n\vert j)W_r^n(\mathcal (D_{j}^{\mathcal X})^c\vert x^n)\leq 2^{-n\tau_6}\right\}
\]
has probability at least $1-2^{-n\tau_6}$.
\end{lem}

As the probability distribution of $\mathcal X$ is not completely standard, we include a proof of this lemma in Appendix \ref{app:transm}, although it does not differ much from the proof in \cite{BBT}. The proof shows that the receiver can even decode the randomization index $l$ in addition to the messages.

\subsubsection{Secrecy}

$\mathcal K^{\mathcal X}_n$ also satisfies the secrecy condition \eqref{eq:permsec} with high probability. Recall that every realization of $\mathcal X$ together with the decoding sets $\{\mathcal D_j^{\mathcal X}:j\in\mathcal J_n\}$ from Lemma \ref{lem:transmission} gives rise to a canonical family of random variables $\mathcal F(\mathcal K_n^{\mathcal X},\mathfrak{\overline W,V})=\{M^n,X^n,Y_r^n,Z_{s^n}^n,\hat M_r^n:r\in\mathcal R,s^n\in\mathcal S^n\}$ as in \eqref{eq:canonnopi}. The dependence of these random variables on $\mathcal X$ is suppressed in the notation. 

\begin{lem}\label{lem:TV-sec}
  For $\delta>0$ sufficiently small, there exist $\tau_1,\tau_2>0$ such that if $n$ is large enough, there exists a family $\{\Theta_{s^n}:s^n\in\mathcal S^n\}$ of finite measures on $\mathcal C^n$ such that the probability of the event 
\[
	\iota_0:=\left\{\max_{j\in\mathcal J_n}\max_{s^n\in\mathcal S^n}\lVert P_{Z^n_{s^n}\vert M^n}(\,\cdot\,\vert j)-\Theta_{s^n}(\cdot)\lVert\leq2^{-\tau_1n}\right\}
\]
is at least $1-2^{-\tau_2n}$. (Note that $P_{Z^n_{s^n}\vert M^n}(\,\cdot\,\vert j)$ is a random variable depending on $\mathcal X$.)
\end{lem}

This lemma is proved in Appendix \ref{app:TV-sec}. 

\begin{cor}\label{cor:TV-sec}
	For $\delta>0$ small enough and $n$ large enough, for the $\tau_1,\tau_2$ from Lemma \ref{lem:TV-sec}, the probability of the event 
\[
	\iota_0':=\left\{\max_{s^n\in\mathcal S^n}I(M^n\wedge Z_{s^n}^n)\leq2^{-\frac{\tau_1}{2}n}\right\}
\]
is at least $1-2^{-\tau_2n}$. (Note again that the joint distribution of $Z^n_{s^n}$ and $M^n$ is a random variable depending on $\mathcal X$.)
\end{cor}

Corollary \ref{cor:TV-sec} immediately follows from Lemma \ref{lem:TV-sec} and the uniform continuity of mutual information in total variation distance \cite[Lemma 2.7]{CK}.

\subsubsection{Synthesis of reliability and secrecy}

Lemma \ref{lem:transmission} and Corollary \ref{cor:TV-sec} show that the probability that $\mathcal K_n^{\mathcal X}$ satisfies \eqref{eq:permrate}-\eqref{eq:permsec} is positive, so a realization satisfying \eqref{eq:permrate}-\eqref{eq:permsec} for $\beta=\tau_1'$ and $R_S=R_S^\dagger(\mathfrak{\overline W,V})-\tau$ must exist. As $\tau>0$ was arbitrary, this proves that $R_S^\dagger(\mathfrak{\overline W,V})$ is an achievable secrecy rate with exponentially decreasing error.

\section{Proof of the achievability part of Theorem \ref{thm:capacity}}

Here we prove that $R_S^*(\mathfrak{W,V})$ is a lower bound to $C^{\max}_{S,\ran}(\mathfrak{W,V})$ and thus by Remark \ref{rem:AVWCKapVergl} also to $C^\mean_{S,\ran}(\mathfrak{W,V})$. We apply the achievability part of Theorem \ref{thm:lowerboundCAVWC} proved in the previous section to a special CAVWC $(\mathfrak{\overline W},\mathfrak V)$. Its determining compound part, the family of stochastic matrices describing communication between the sender and the legitimate receiver, is given by $\{W_q:q\in\mathcal P(\mathcal S)\}$, where $W_q:=\sum_{s\in\mathcal S}W_sq(s)$. We thus obtain $\mathfrak{\overline W}=\{W_q^n:q\in\mathcal P(\mathcal S), n=1,2,\ldots\}$. Observe that for $R_S^*(\mathfrak{W,V})$ defined in \eqref{eq:R_S}, we have
\[
  R_S^*(\mathfrak{W,V})=R_S^*(\mathfrak{\overline W,V}).
\]
Central to the proof is Ahlswede's robustification technique:

\begin{lem}[\cite{A3}]\label{lem:RT}
  If a function $f:\mathcal S^n\rightarrow [0,1]$ satisfies
\begin{equation}\label{eq:condRT}
  \sum_{s^n\in S^n}f(s^n)q(s_1)\cdots q(s_n)\geq 1-\varepsilon
\end{equation}
for all $q\in\mathcal P_0^n(\mathcal S)$ and some $\varepsilon\in[0,1]$, then
\begin{equation}\label{eq:conclRT}
  \frac{1}{n!}\sum_{\pi\in\Pi_n}f(\pi(s^n))\geq 1-3\cdot(n+1)^{\lvert\mathcal S\rvert}\cdot\varepsilon.
\end{equation}
\end{lem}

Let now $\varepsilon>0$. By Theorem \ref{thm:lowerboundCAVWC} applied to the CAVWC $(\mathfrak{\overline W,V})$ defined above, there exists a $\beta>0$ such that for sufficiently large $n$, there exists an uncorrelated $(n,J_n)$-code $\mathcal K_n$ satisfying
\begin{align}
	\frac{1}{n}\log J_n\geq R_S^*(\mathfrak{\overline W,V})-\varepsilon=R_S^*(\mathfrak{W,V})-\varepsilon,\notag\\
  \bar e(\mathcal K_n)=\max_{q\in\mathcal P(\mathcal S)}\frac{1}{J_n}\sum_{j=1}^{J_n}\sum_{x^n\in\mathcal A^n}E(x^n\vert j)W_q^n(\mathcal D_j^c\vert x^n)\leq 2^{-n(\beta-\varepsilon)},\label{eq:robuerr}\\  
  \max_{s^n\in S^n}\max_{\pi\in\mathcal S_n}I(M^n\wedge Z_{s^n}^n(\pi))\leq\varepsilon.\label{eq:robusec}
\end{align}

Define the function $f$ by
\[
  f(s^n):=\frac{1}{J_n}\sum_{j\in\mathcal J_n}\sum_{x^n\in\mathcal A^n}E(x^n\vert j)W_{s^n}^n(\mathcal D_j\vert x^n).
\]
It was already noted in Remark \ref{rem:remCAVWC}-4) that for any $q\in\mathcal P_0^n(\mathcal S)$ and $x^n\in\mathcal A^n$ and $y^n\in\mathcal B^n$
\[
  \sum_{s^n}W_{s^n}^n(y^n\vert x^n)q(s_1)\cdots q(s_n)=W_q^n(y^n\vert x^n).
\]
Thus by \eqref{eq:robuerr}
\begin{align*}
  \sum_{s^n\in\mathcal S^n}f(s^n)q(s_1)\cdots q(s_n)&
  =\frac{1}{J_n}\sum_{j\in\mathcal J_n}\sum_{s^n\in\mathcal S^n}\sum_{x^n\in\mathcal A^n}E(x^n\vert j)W_{s^n}^n(\mathcal D_j\vert x^n)q(s_1)\cdots q(s_n)\\
  &=\frac{1}{J_n}\sum_{j\in\mathcal J_n}\sum_{s^n\in\mathcal S^n}\sum_{x^n\in\mathcal A^n}E(x^n\vert j)W_{q}^n(\mathcal D_j\vert x^n)\\
  &\geq 1-2^{-n(\beta-\varepsilon)}.
\end{align*}

Now we derive a correlated random $(n,J_n)$-code $\mathcal K_n^\ran$ from $\mathcal K_n$. Let $E^\pi$ be given by $E^\pi(x^n\vert j):=E(\pi^{-1}(x^n)\vert j)$ and let $\mathcal D_j^\pi:=\{\pi(y^n):y^n\in\mathcal D_j\}$. Further let $G_n$ be uniformly distributed on this family indexed by $\Pi_n$. One has
\begin{align*}
  1-e(\mathcal K_n^\ran)&=\frac{1}{n!}\sum_{\pi\in\Pi_n}\frac{1}{J_n}\sum_{j\in\mathcal J_n}\sum_{x^n}E^{\pi^{-1}}(x^n\vert j)W_{s^n}^n(\mathcal D_j^{\pi^{-1}}\vert x^n)\\
  &=\frac{1}{n!}\sum_{\pi\in\Pi_n}\frac{1}{J_n}\sum_{j\in\mathcal J_n}\sum_{x^n}E(\pi(x^n)\vert j)W_{s^n}^n(\mathcal D_j^{\pi^{-1}}\vert x^n)\\
  &=\frac{1}{n!}\sum_{\pi\in\Pi_n}\frac{1}{J_n}\sum_{j\in\mathcal J_n}\sum_{x^n}E(x^n\vert j)W_{s^n}^n(\mathcal D_j^{\pi^{-1}}\vert\pi^{-1}(x^n))\\
  &=\frac{1}{n!}\sum_{\pi\in\Pi_n}\frac{1}{J_n}\sum_{j\in\mathcal J_n}\sum_{x^n}E(x^n\vert j)W_{\pi(s^n)}^n(\mathcal D_j\vert x^n).
\end{align*}
With $\varepsilon=2^{-n(\beta-\varepsilon)}$, Lemma \ref{lem:RT} implies that the last term is lower-bounded by $1-(n+1)^{\lvert S\rvert}2^{-n(\beta-\varepsilon)}\geq 1-2^{-n(\beta-2\varepsilon)}$ for sufficiently large $n$. This settles the reliability properties of $\mathcal K_n^\ran$.

The secrecy properties of $\mathcal K_n^\ran$ are immediate, as \eqref{eq:robusec} implies
\[
  \frac{1}{n!}\sum_{\pi\in\Pi_n}I(M^n\wedge Z_{s^n}^n(\pi))\leq\max_{\pi\in\Pi_n}I(M^n\wedge Z_{s^n}^n(\pi))\leq\varepsilon
\]
for every $s^n\in\mathcal S^n$. Hence $R_S^*(\mathfrak{\overline W,V})$ is an achievable correlated random coding maximal secrecy rate.

\section{The Converses}\label{sect:conv}

One unusual difficulty arises in the proof of the converse of Theorem \ref{thm:capacity}. This difficulty consists in the fact that the common randomness prohibits a ``naive'' application of the data processing inequality. It is thus necessary to limit the amount of common randomness of an arbitrary correlated random code in order to overcome this difficulty. This has already been done in Lemma \ref{lem:CRreduction}.

Let $R_S<C_{S,\ran}^\mean(\mathfrak{W,V})$. From Lemma \ref{lem:CRreduction} we know that for every $\varepsilon>0$ there is an $L=L(R_S,\varepsilon)$ such that for sufficiently large $n$ there is a correlated random $(n,J_n)$-code $\mathcal K_n^\ran$ satisfying
\begin{align}
	\frac{1}{n}\log J_n&\geq R_S-\varepsilon,\label{eq:conv0}\\
  e(\mathcal K_n^\ran)&\leq\varepsilon,\label{eq:conv1}\\
  \max_{s^n\in\mathcal S^n}I(M^n\wedge Z_{s^n}^n\vert G_n)&\leq\varepsilon\label{eq:conv2},\\
  \lvert\supp(G_n)\rvert&=L(R_S,\varepsilon).
\end{align}

By \cite[Lemma 12.3]{CK}, the average error incurred by any uncorrelated code $\mathcal K_n$ used over the AVC $\mathfrak W$ equals the average error of $\mathcal K_n$ over the AVC determined by the convex hull of $\{W_s:s\in\mathcal S\}$, i.~e.\ the AVC $\{W_{q^n}^n:q^n\in\mathcal P(\mathcal S)^n,n=1,2,\ldots\}$, where
\[
  W_{q^n}^n(y^n\vert x^n):=\prod_{i=1}^n\sum_{s_i\in\mathcal S}W_{s_i}(y_i\vert x_i)q_i(s_i).
\]
This is a simple consequence of the fact that the average error is affine in the channel and carries over to correlated random codes. Hence \eqref{eq:conv1} implies
\begin{equation}\label{eq:fehlerkonvex}
  \max_{q^n\in\mathcal P(\mathcal S)^n}\frac{1}{J_n}\sum_{j=1}^{J_n}\sum_{\gamma\in\Gamma_n}\sum_{x^n\in\mathcal A^n}E^\gamma(x^n\vert j)W_{q^n}^n\bigl((\mathcal D_j^\gamma)^c\vert x^n\bigr)P_{G_n}(\gamma)\leq\varepsilon.
\end{equation}
From \eqref{eq:fehlerkonvex}, one infers that the average error of $\mathcal K_n^\ran$ for transmission over the compound channel $\mathfrak{\overline W}$ is upper-bounded by $\varepsilon$ as well, i.~e.\
\begin{equation}\label{eq:fehlercomp}
  \max_{q\in\mathcal P(\mathcal S)}\frac{1}{J_n}\sum_{j=1}^{J_n}\sum_{\gamma\in\Gamma_n}\sum_{x^n\in\mathcal A^n}E^\gamma(x^n\vert j)W_{q}^n\bigl((\mathcal D_j^\gamma)^c\vert x^n\bigr)P_{G_n}(\gamma)\leq\varepsilon.
\end{equation}

Due to Fano's inequality \cite[Lemma 3.8]{CK}, \eqref{eq:fehlercomp} implies for every $q\in\mathcal P(\mathcal S)$
\begin{align*}
  H(M^n\vert\hat M^n_q,G_n)&=\sum_{\gamma\in\supp(G_n)}H(M^n\vert\hat M_q^n,G_n=\gamma)P_{G_n}(\gamma)\\
  &\leq1+\sum_{\gamma\in\supp(G_n)}\mathbb P[M^n\neq\hat M^n_q\vert G_n=\gamma]P_{G_n}(\gamma)\log J_n\\
  &=1+\varepsilon\log J_n.
\end{align*}
Here the $\hat M_q^n$ are the random variables from the canonical family $\mathcal F(\mathcal K_n^\ran,\mathfrak{\overline W,V})$ defined in \eqref{eq:canonCAVWC}. Hence the independence of $M^n$ and $G_n$ yields
\[
	\log J_n=H(M^n)=H(M^n\vert G_n)\\=I(M^n\wedge\hat M^n_q\vert G_n)+H(M^n\vert\hat M^n_q,G_n)\leq I(M^n\wedge\hat M^n_q\vert G_n)+1+\varepsilon\log J_n,
\]
so by rearranging and taking \eqref{eq:conv2} into account, we have for every $q\in\mathcal P(\mathcal S)$ and $s^n\in\mathcal S^n$
\[
  (1-\varepsilon)\log J_n\leq I(M^n\wedge\hat M^n_q\vert G_n)-I(M^n\wedge Z_{s^n}^n\vert G_n)+1+\varepsilon.
\]

We have to get rid of $G_n$ in some way. The only reasonable way to achieve this seems to be through the use of the convexity of the mutual information in the channel argument. But while this is a valid choice for the ``secrecy term'', it is certainly invalid for the ``legal'' term. This is due to the fact that $G_n$ is independent of $M^n$, but not of $\hat M_q^n$ or $Y_q^n$. An application of the data processing inequality is thus only possible conditioned on $G_n$. It is here where the importance of Lemma \ref{lem:CRreduction} becomes evident: The cardinality of the support of $G_n$ is bounded and independent of $n$ for $n$ sufficiently large, hence we can write
\begin{align*}
	I(M^n\wedge\hat M^n_q\vert G_n)&=H(M^n)-H(M^n\vert Y_q^n,G_n)\\
	&\leq H(M^n)-H(M^n\vert Y_q^n)+H(G_n)\\
	&\leq I(M^n\wedge Y_q^n)+\log L(R_S,\varepsilon),
\end{align*}
where we employed the fact that $H(S)\leq H(S,T)=H(S\vert T)+H(T)$. Thus if $n$ is sufficiently large, we obtain that
\begin{align}
	\frac{1}{n}\log J_n&\leq\frac{1}{n(1-\varepsilon)}\bigl(\min_{q\in\mathcal P(\mathcal S)}I(M^n\wedge\hat M_q^n\vert G_n)-\max_{s^n\in\mathcal S^n}(M^n\wedge Z_{s^n}^n\vert G_n)+1+\varepsilon\bigr)\notag\\
	&\leq\frac{1}{n(1-\varepsilon)}\bigl(\min_{q\in\mathcal P(\mathcal S)}I(M^n\wedge Y_q^n)-\max_{s^n\in\mathcal S^n}I(M^n\wedge Z_{s^n}^n)\bigr)+\frac{\log L(R_S,\varepsilon)+1+\varepsilon}{n(1-\varepsilon)}.\label{eq:logL}
\end{align}
For $n$ sufficiently large, as $L(R_S,\varepsilon)$ is independent of $n$, the second term of \eqref{eq:logL} is upper-bounded by $\varepsilon$. If we set $\bar U:=M^n$ and $\bar X^n:=X^n$ and $\bar Y_q^n:=Y_q^n$ and $\bar Z_{s^n}^n:=Z_{s^n}^n$, the joint distributions
\begin{align*}
	P_{\bar U\bar X^n\bar Y_q^n}(j,x^n,y^n)=\frac{1}{J_n}\sum_{\gamma\in\Gamma_n}P_{G_n}(\gamma)E^\gamma(x^n\vert j)W_q^n(y^n\vert x^n),\\
	P_{\bar U\bar X^n\bar Z_{s^n}^n}(j,x^n,z^n)=\frac{1}{J_n}\sum_{\gamma\in\Gamma_n}P_{G_n}(\gamma)E^\gamma(x^n\vert j)V_{s^n}^n(z^n\vert x^n)
\end{align*}
have the form required in the definition of $R_S^*(\mathfrak{W,V})$, and the shared randomness is now completely reduced to randomness at the encoder. Thus by \eqref{eq:conv0} and as $\varepsilon$ was arbitrary, we have $R_S\leq R_S^*(\mathfrak{W,V})$, hence $C_{S,\ran}^\mean(\mathfrak{W,V})\leq R_S^*(\mathfrak{W,V})$, and therefore also $C_{S,\ran}^{\max}(\mathfrak{W,V})\leq R_S^*(\mathfrak{W,V})$. This completes the proof of the converse of Theorem \ref{thm:capacity}.

\begin{rem}\label{rem:zusatz-conv}
	As the average error is affine in the channel, one can even pass to a maximum over $\tilde q\in\mathcal P(\mathcal S^n)$ in \eqref{eq:fehlerkonvex}. Skipping the reduction to $q\in\mathcal P(\mathcal S)$ in \eqref{eq:fehlercomp} and directly applying Fano's inequality, the rest of the proof can be performed as above for every $\tilde q\in\mathcal P(\mathcal S^n)$ using random variables $Y_{\tilde q}^n=\bar Y_{\tilde q}^n$ defined by 
	\[
		P_{Y_{\tilde q}^n\vert X^n}(y^n\vert x^n)=\sum_{s^n}\tilde q(s^n)W^n_{s^n}(y^n\vert x^n).
	\]
	This shows that the right-hand side of \eqref{eq:ersteabw} upper-bounds $C_{ S,\ran}^\mean(\mathfrak{W,V})=R_S^*(\mathfrak{W,V})$. Since the right-hand side of \eqref{eq:ersteabw} trivially is a lower bound on $R_S^*(\mathfrak{W,V})$, as noted in Remark \ref{rem:remCAVWC}-\ref{rem:remCAVWC-zusatz}), we can conclude the validity of equality \eqref{eq:ersteabw}.
\end{rem}

The converse for Theorem \ref{thm:lowerboundCAVWC} follows the same lines. It is simpler as no common randomness has to be considered.

\section{Discussion}\label{sect:disc}

The main result of this paper is the correlated random coding secrecy capacity of the AVWC for the case where  the eavesdropper is allowed access to the correlated randomness shared by sender and intended receiver. Applying Ahlswede's robustification technique, the main problem was solved via reduction to the secrecy capacity problem of the CAVWC, which is compound between the sender and the intended receiver and arbitrarily varying between the sender and the eavesdropper.  

The secrecy capacity formula obtained in the main theorem is a multi-letter formula. Of course, this makes a direct computation impossible. On the other hand, it is not known whether a general, computable, single-letter formula exists at all. For a given AVWC, the value of the multi-letter formula can be approximated by restricting computation to a finite number of letters. An open problem not addressed in this paper is the goodness of finite-letter approximation.

However, the use of a capacity formula is much larger than just to calculate the capacity. It can be applied in the in-depth analysis of the channels in question. For example, using nothing but the capacity formula, it can be shown for discrete memoryless channels that the capacity of parallel channels is the sum of their capacities. For the AVWC, an analysis of the capacity formula shows that the correlated random coding secrecy capacity is continuous in the AVWC, which is impossible to derive a priori. This result is of great engineering importance because it ensures that small variations in the channel data cannot lead to completely different secrecy capacities. This is very reassuring, as lots of resources would otherwise have to be spent on channel estimation. In fact, the necessary precision of the channel estimate would grow without limits the closer the channel would be to a point of discontinuity of the secrecy capacity function.

Follow-up work on the AVWC correlated random coding secrecy capacity for the case that the eavesdropper has no knowledge of the correlated randomness as well as the AVWC uncorrelated coding secrecy capacity is presented in \cite{AVWCII}.

\appendices

\section{Proof of Corollary \ref{cor:MolavianJazi}}\label{app:MolavianJazi}

It is obvious that the right-hand side of \eqref{eq:MolavianJazi} is upper-bounded by $R_S^*(\mathfrak{W,V})$, see Remark \ref{rem:remCAVWC}-\ref{rem:remCAVWC-limsup}). Thus it remains to show the converse relation. Let $k$ be a positive integer and let $\{\bar U,\bar X,\bar Y_{q_1}^k,\bar Z_{s^k}^k\}$ be a family of random variables as in the definition of $R_S^*(\mathfrak{W,V})$. The existence of a best channel to the eavesdropper guarantees that $I(U\wedge Z_{s_2^k}^k)\leq I(U\wedge Z_{s_*}^k)$ for every $s_2^k\in\mathcal S_2^k$, where $P_{\bar Z_{s_*}^k\vert\bar X}(z^k\vert x^k)=\prod_{i=1}^kV_{s_*}(z_i\vert x_i)$. In particular, $I(U\wedge Z_{s_*}^k)=\max_{s_2\in\mathcal S_2}I(U\wedge Z_{s_2^k}^k)$. Therefore
\begin{align}
	\frac{1}{k}\biggl(\min_{q_1\in\mathcal P(\mathcal S_1)}I(\bar U\wedge\bar Y_{q_1}^k)-\max_{s_2^k\in\mathcal S_2^k}I(\bar U\wedge\bar Z_{s^k}^k)\biggr)
	&=\frac{1}{k}\min_{q_1\in\mathcal P(\mathcal S_1)}\biggl(I(\bar U\wedge\bar Y_{q_1}^k)-I(\bar U\wedge\bar Z_{s_*}^k)\biggr)\notag\\
	&\leq\frac{1}{k}\min_{q_1\in\mathcal P(\mathcal S_1)}I(\bar U\wedge\bar Y_{q_1}^k\vert\bar Z_{s_*}^k), \label{eq:MBL1}
\end{align}
where strong degradedness was applied in \eqref{eq:MBL1}. In a similar fashion as in the derivation of (23)-(26) in \cite{MBL}, one can rewrite the right-hand side of \eqref{eq:MBL1} as $I(\bar X^*\wedge\bar Y_{q_1}^*\vert\bar Z_{s_*}^*)$, where $\bar X^*$ is a random variable on $\mathcal A$ and the distributions of $\bar Y_{q_1}^*$ and $\bar Z_{s_*}^*$ satisfy $P_{\bar Y_{q_1}^*\vert\bar X^*}=W_{q_1}$ and $P_{\bar Z_{s_*}^*\vert\bar X^*}=V_{s_*}$. Again using the strong degradedness of $(\mathfrak{W,V})$ and the existence of a best channel to the eavesdropper and defining $\bar Z_{s_2}^*$ by its conditional distribution $P_{\bar Z_{s_2}^*\vert\bar X^*}=V_{s_2}$ for every $s_2\in\mathcal S_2$, one obtains
\begin{align*}
	\min_{q_1\in\mathcal P(\mathcal S_1)}I(\bar X^*\wedge\bar Y_{q_1}^*\vert\bar Z_{s_*}^*)
	&\leq\min_{q_1\in\mathcal P(\mathcal S_1)}\biggl(I(\bar X^*\wedge\bar Y_{q_1}^*)-I(\bar X^*\wedge\bar Z_{s_*}^*)\biggr)\\
	&=\min_{q_1\in\mathcal P(\mathcal S_1)}I(\bar X^*\wedge\bar Y_{q_1}^*)-\max_{s_2\in\mathcal S_2}I(\bar X^*\wedge\bar Z_{s_*}^*).
\end{align*}

Inserting this in the definition of $R_S^*(\mathfrak{W,V})$ shows that $R_S^*(\mathfrak{W,V})$ is upper-bounded by the right-hand side of \eqref{eq:MolavianJazi}, thus proving that \eqref{eq:MolavianJazi} indeed is an equality. This proves Corollary \ref{cor:MolavianJazi}.

\section{Proof of Lemma \ref{lem:permutations}}\label{app:permutations}

Assume $\mathcal K_n$ satisfies \eqref{eq:gegeben} and has stochastic encoder $E$. Recall that $E^\pi$ is defined by $E^\pi(x^n\vert j):=E(\pi^{-1}(x^n)\vert j)$. The random variables below are from the canonical permutation-invariant family $\mathcal F(\mathcal K_n,\mathfrak{W,V},\Pi_n)$.

\begin{lem}\label{lem:permdist}
	For every $\pi\in\Pi_n$, we have $P_{M_n\pi(Z_{s^n}^n(\id))}=P_{M_nZ_{\pi(s^n)}^n(\pi)}$.
\end{lem}

\begin{IEEEproof}
	Let $j\in\mathcal J_n$ and $z^n\in\mathcal C^n$. Then
	\begin{align*}
		\mathbb P[M_n=j,\pi(Z_{s^n}^n(\id))=z^n]&=\mathbb P[M_n=j,Z_{s^n}^n(\id)=\pi^{-1}(z^n)]\\
		&=\frac{1}{J_n}\sum_{x^n}E(x^n\vert j)V_{s^n}^n(\pi^{-1}(z^n)\vert x^n)\\
		&=\frac{1}{J_n}\sum_{x^n}E(\pi^{-1}(x^n)\vert j)V_{s^n}^n(\pi^{-1}(z^n)\vert\pi^{-1}(x^n))\\
		&=\frac{1}{J_n}\sum_{x^n}E^\pi(x^n\vert j)V_{\pi(s^n)}^n(z^n\vert x^n)\\
		&=\mathbb P[M_n=j,Z_{\pi(s^n)}^n(\pi)=z^n].
	\end{align*}
\end{IEEEproof}

Now assume that \eqref{eq:gegeben} holds. Then
\begin{align*}
	\max_{\pi\in\Pi_{n}}\max_{s^{n}}I(M_n\wedge Z_{s^{n}}^{n}(\pi))
	&=\max_{\pi\in\Pi_{n}}\max_{s^{n}}I(M_n\wedge Z_{\pi(s^{n})}^{n}(\pi))\\
	&\stackrel{(i)}{=}\max_{\pi\in\Pi_{n}}\max_{s^{n}}I(M_n\wedge\pi(Z_{s^{n}}^{n}(\id)))\\
	&\stackrel{(ii)}{\leq}\max_{s^n}I(M_n\wedge Z_{s^n}^n(\id))\\
	&\leq\varepsilon
\end{align*}
where Lemma \ref{lem:permdist} was applied in (i) and the data processing inequality in (ii). Thus \eqref{eq:gegeben} implies \eqref{eq:gewollt}.

\section{Channel prefixing}\label{app:prefixrand}

Assume for any CAVWC $(\tilde{\mathfrak{\overline W},\tilde{\mathfrak V}})$ that $R_S^\dagger(\tilde{\mathfrak{\overline W}},\tilde{V})$ is achievable with exponentially decreasing error for $(\tilde{\mathfrak{\overline W}},\tilde{V})$. We have to show that then for a given CAVWC $(\mathfrak{\overline W,V})$, $R_S^*(\mathfrak{\overline W,V})$ also is an achievable rate with exponentially decreasing error for $(\mathfrak{\overline W,V})$. Choose a positive integer $k$, a finite subset $\mathcal U$ of the integers, and a stochastic matrix $T:\mathcal U\rightarrow\mathcal P(\mathcal A^k)$. For every $r\in\mathcal R$ and $s^k\in\mathcal S^k$, this induces stochastic matrices $\tilde W_r:\mathcal U\rightarrow\mathcal P(\mathcal B^k)$ and $\tilde V_{s^k}:\mathcal U\rightarrow\mathcal P(\mathcal C^k)$ defined by
\begin{align*}
	\tilde W_r(y^k\vert u)&:=\sum_{x^k}T(x^k\vert u)W_r^k(y^k\vert x^k),\\
	\tilde V_{s^k}(y^k\vert u)&:=\sum_{x^k}T(x^k\vert u)V_{s^k}^k(z^k\vert x^k).
\end{align*} 
This induces families
\begin{align*}
	\tilde{\overline{\mathfrak{W}}}&:=\{\tilde W_r^n:r\in\mathcal R,n=1,2,\ldots\},\\
	\tilde{\mathfrak{V}}&:=\{\tilde V_{s^{kn}}^n:s^{kn}\in(\mathcal S^k)^n,n=1,2,\ldots\},
\end{align*}
and hence a CAVWC denoted by $(\tilde{\overline{\mathfrak W}},\tilde{\mathfrak V})$. The compound part of this channel also has $\mathcal R$ as its state set, the state set of the eavesdropper channel equals $\mathcal S^k$. By assumption, $R_S^\dagger(\tilde{\overline{\mathfrak W}},\tilde{\mathfrak V})$
is an achievable rate with exponentially decreasing error for $(\tilde{\overline{\mathfrak W}},\tilde{\mathfrak V})$. Thus there exists a $\beta>0$ such that for every $\varepsilon>0$ and sufficiently large $n$, one obtains an $(n,J_n)$-code $\tilde{\mathcal K}_n$ for $(\tilde{\overline{\mathfrak W}},\tilde{\mathfrak V})$ with canonical random family $\mathcal F(\tilde{\mathcal{K}}_n,\tilde{\overline{\mathfrak W}},\tilde{\mathfrak V})=\{\tilde M^n,\tilde U^n,\tilde Y_r^{kn},\tilde Z_{s^{kn}}^{kn},\tilde{\hat M}^n_r:r\in\mathcal R,s^{kn}\in(\mathcal S^k)^n\}$ satisfying
\begin{align}
  \frac{1}{n}\log J_n&\geq R_S^\dagger(\tilde{\overline{\mathfrak W}},\tilde{\mathfrak V})-\varepsilon,\\
  -\frac{1}{n}\log\bar e(\tilde{\mathcal{K}}_n)&\geq\beta-\varepsilon,\\
	\max_{s^{kn}\in(\mathcal S^k)^n}I(\tilde M^n\wedge\tilde Z_{s^{kn}}^{kn})&\leq\varepsilon.
\end{align}
Now define the stochastic encoder $E:\mathcal J_n\rightarrow\mathcal P(\mathcal A^{kn})$ through
\[
  E(x^{kn}\vert j):=\sum_{u^n\in\mathcal U^n}E^*(u^n\vert j)T^n(x^{kn}\vert u^n).
\]
Together with the decoding sets $\mathcal D_j^*$ considered as sets $\mathcal D_j\subset\mathcal B^{kn}$, this defines an uncorrelated $(kn,J_n)$-code $\mathcal K_{kn}$ for the CAVWC $(\mathfrak{\overline W,V})$. Observe that, if $\mathcal F(\mathcal K_{kn},\mathfrak{\overline W,V})=\{M^n,X^n,Y_r^{kn},Z_{s^{kn}}^{kn},M^n_r:r\in\mathcal R,s^{kn}\in\mathcal S^{kn}\}$ is the canonical random family of $\mathcal K_{kn}$, then for every $r\in\mathcal R_n$ and $s^{kn}$ regarded either as en element of $\mathcal S^{kn}$ or $(\mathcal S^k)^n$, the joint probability of $(M^n,Y_r^{kn},Z_{s^{kn}}^{kn},\hat M_r^{kn})$ equals that of $(\tilde M^n,\tilde Y_r^{kn},\tilde Z_{s^{kn}}^{kn},\tilde{\hat M}_r^n)$.

It immediately follows that
\begin{align*}
	\frac{1}{kn}\log J_n&\geq\frac{1}{k} R_S^\dagger(\tilde{\overline{\mathfrak{W}}},\tilde{\mathfrak V})-\frac{\varepsilon}{k},\\
	- \frac{1}{kn}\bar e(\mathcal K_{kn})&\geq \frac{\beta-\varepsilon}{k},\\
	\max_{s^{kn}\in\mathcal S^{kn}}I(M^n\wedge Z_{s^{kn}}^{kn})&\leq\varepsilon.
\end{align*}
Thus after optimization over $T$ and $k$, it follows that $R_S^*(\mathfrak{\overline W,V})$ is an achievable secrecy rate with exponentially decreasing error for $(\mathfrak{\overline W,V})$.

\section{Types and typical sequences}\label{app:typesandtypical}

The proofs require some facts about types and typical sequences. For reference, we include them here. $\mathcal A,\mathcal B$ and $W,\tilde W$ are generic sets/stochastic matrices.

\begin{lem}\label{lem:wknf}
  Let $\bar X$ be an $\mathcal A$-valued random variable and let $x^n\in\mathcal T_{\bar X,\delta}^n$. Further let $W:\mathcal A\longrightarrow\mathcal P(\mathcal S)$. Then for any $\mathcal B$-valued random variable $\bar Y$ with $P_{\bar Y\vert\bar X}=W$,
\begin{align*}
  \lvert\mathcal T_{\bar Y,\delta}^n\rvert&\leq\exp\{n(H(\bar Y)+f_1(\delta))\},\\
  W^n(y^n\vert x^n)&\leq\exp\{-n(H(\bar Y\vert\bar X)-f_2(\delta))\}\quad\text{for all }y^n\in\mathcal T_{\bar Y\vert\bar X,\delta}^n(x^n)
\end{align*}
with universal $f_1(\delta),f_2(\delta)>0$ satisfying $\lim_{\delta\rightarrow0}f_1(\delta)=\lim_{\delta\rightarrow0}f_2(\delta)=0$.
\end{lem}

\begin{lem}\label{lem:AEP}
  Let $\delta>0$. Let $(\bar X,\bar Y)$ assume values in $\mathcal A\times\mathcal B$ such that $P_{\bar Y\vert\bar X}=W$, for some $W:\mathcal A\longrightarrow\mathcal P(\mathcal B)$, and let $x^n\in\mathcal A^n$. There exist a universal $c'>0$ and an $n_0=n_0(\lvert\mathcal A\rvert,\lvert\mathcal B\rvert,\delta)\geq 1$ such that for $n\geq n_0$
\begin{align*}
  P_{\bar X}^n(\mathcal T_{\bar X,\delta}^n)&\geq 1-2^{-nc'\delta^2},\\
  W^n(\mathcal T_{\bar Y\vert\bar X,\delta}^n(x^n)\vert x^n)&\geq 1-2^{-nc'\delta^2}.
\end{align*}
\end{lem}

\begin{lem}\label{lem:cardtyp}
  The cardinality of $\mathcal P_0^n(\mathcal S)$ is upper-bounded by $(n+1)^{\lvert\mathcal S\rvert}$.
\end{lem}

The proofs of Lemmas \ref{lem:wknf}-\ref{lem:cardtyp} can be found in e.g.\ \cite{CK}. A proof of the next lemma can be found in \cite{Bjelakovic2013}.

\begin{lem}\label{lem:verschtypen}
  Let $(\bar X,\bar Y)$ and $(\bar X',\bar Y')$ two pairs of $\mathcal A\times\mathcal B$-valued random variables. Then for sufficiently small $\delta>0$ and any positive integer $n$,
\begin{equation}\label{eq:hs-o}
  P_{\bar Y}^n(\mathcal T_{\bar Y'\vert\bar X',\delta}^n(x^n))\leq (n+1)^{\lvert\mathcal A\rvert\lvert\mathcal B\rvert}\exp\{-n(I(\bar X'\wedge\bar Y')-f_3(\delta))\}
\end{equation}
for all $\tilde x^n\in\mathcal T_{\bar X',\delta}^n$ holds for a universal $f_3(\delta)>0$ with $\lim_{n\rightarrow\infty}f_3(\delta)=0$.
\end{lem}

Note that the right-hand side of \eqref{eq:hs-o} does not depend on $(\bar X,\bar Y)$, so one might wonder how sharp this bound is. But we will apply the lemma in a case where $\bar X=\bar X'$ and where $P_{\bar Y\vert\bar X}$ and $P_{\bar Y'\vert\bar X'}$ may be close (see Appendix \ref{app:transm}). Thus it turns out to give the correct upper bound.

\section{Proof of Lemma \ref{lem:transmission}}\label{app:transm}
The fact that the probability of $\bar e(\mathcal K_n^{\mathcal X})$ being small is large is well-known in principle, cf.\ \cite{CK}. As our choice of codewords does not quite follow the standard approach and we use stochastic encoders, we present the proof nonetheless. We start with a lemma which assumes a finite state set for $\mathfrak{\overline W}$ and actually shows that the sender can also reliably decode the randomization index with high probability.

\begin{lem}\label{lem:transmissionhelp}
  Let $\mathcal R'\subset\mathcal R$ be finite. With
\[
  \mathcal{\hat D}_{jl}^{\mathcal X}:=\bigcup_{r\in\mathcal R}\mathcal T_{\bar Y_r\vert\bar X,\delta}^n(X_{jl}),
\]
define
\[
  \tilde{\mathcal D}_{jl}^{\mathcal X}:=\mathcal{\hat D}_{jl}^{\mathcal X}\cap\Bigl(\bigcup_{(j',l')\in\mathcal J_n\times\mathcal L_n\setminus\{(j,l)\}}\mathcal{\hat D}_{j'l'}^{\mathcal X}\Bigr)^c.
\]
In order for these decoding sets to cover the complete output space, we assume without loss of generality that $\tilde{\mathcal D}_{11}^{\mathcal X}$ contains all $y^n\in\mathcal B^n$ not assigned to any message so far. This does not increase the average error. The $\tilde{\mathcal D}_{jl}^{\mathcal X}$ are pairwise disjoint ($(j,l)\in\mathcal J_n\times\mathcal L_n$). For $\tau\geq\tau_0(\delta)$, with $\tau_0(\delta)\rightarrow0$ as $\delta>0$, there exists an $a=a(\tau,\delta)>0$ such that the event
\[
  \tilde\iota_3:=\left\{\max_{r\in\mathcal R'}\frac{1}{J_nL_n}\sum_{(j,l)\in\mathcal J_n\times\mathcal L_n}W_r^n\bigl((\tilde{\mathcal D}_{jl}^{\mathcal X})^c\vert X_{jl}\bigr)\leq 2^{-na}\right\}
\]
has probability at least $1-2^{-na}$.
\end{lem}

\begin{IEEEproof}
  The disjointness of the decoding sets is obvious. We first show an upper bound on the mean error incurred by $\mathcal K_n^{\mathcal X}$ for given state $r\in\mathcal R'$. More precisely, setting
\[
  e_r(\mathcal K_n^{\mathcal X}):=\frac{1}{J_nL_n}\sum_{j=1}^{J_n}\sum_{l=1}^{L_n}W^n_r((\tilde{\mathcal D}_{jl}^{\mathcal X})^c\vert X_{jl}),
\]
we claim
\begin{equation}\label{eq:errorgivenq}
  \mathbb E\left[e_r(\mathcal K_n^{\mathcal X})\right]\leq 2^{-na'}
\end{equation}
for some $a'=a'(\tau,\delta)>0$ and for $n$ sufficiently large. The left-hand side of \eqref{eq:errorgivenq} equals
\begin{align}
  &\mathbb E\left[W^n_r((\tilde{\mathcal D}_{11}^{\mathcal X})^c\vert X_{11})\right]\notag\\
  &\leq\mathbb E\left[W^n_r((\mathcal{\hat D}_{11}^{\mathcal X})^c\vert X_{11})\right]\label{eq:coding3}\\
  &+\sum_{\substack{(j,l)\in\mathcal J_n\times\mathcal L_n:\\(j,l)\neq (1,1)}}\mathbb E\left[W^n_r(\mathcal{\hat D}_{jl}^{\mathcal X}\vert X_{11})\right].\label{eq:coding1}
\end{align}
For \eqref{eq:coding3}, we have
\[
  \mathbb E\left[W^n_r((\mathcal{\hat D}_{11}^{\mathcal X})^c\vert X_{11})\right]\leq\mathbb E\left[W^n_r((\mathcal T_{\bar Y_r\vert\bar X,\delta}^n(X_{11}))^c\vert X_{11})\right],
\]
which by Lemma \ref{lem:AEP} is upper-bounded by $2^{-nc'\delta^2}$ for $n$ sufficiently large. Thus \eqref{eq:coding3} is upper-bounded by the same number. For each of the terms in \eqref{eq:coding1}, we obtain
\begin{align*}
  \mathbb E\left[W^n_r(\mathcal{\hat D}_{jl}^{\mathcal X}\vert X_{11})\right]\leq\sum_{r'\in\mathcal R'}\mathbb E\left[W_r^n(\mathcal T_{\bar Y_{r'}\vert\bar X,\delta}^n(X_{jl})\vert X_{11})\right].
\end{align*}
For sufficiently large $n$, the terms on the right-hand side can be written (recall that $(j,l)\neq (1,1)$)
\begin{align}
  &\mathbb E\left[W_r^n(\mathcal T_{\bar Y_{r'}\vert\bar X,\delta}^n(X_{jl})\vert X_{11})\right]\notag\\
  &=\sum_{x^n,\tilde x^n\in\mathcal T_{\bar X,\delta}^n}W_r^n(\mathcal T_{\bar Y_{r'}\vert\bar X,\delta}^n(\tilde x^n)\vert x^n)P'(x^n)P'(\tilde x^n)\notag\\
  &\stackrel{(i)}{\leq}(1-2^{-nc'\delta})^{-2}\sum_{\tilde x^n\in T_{\bar X,\delta}^n}P_{\bar Y_r}^n(\mathcal T_{\bar Y_{r'}\vert\bar X,\delta}^n(\tilde x^n))P_{\bar X}^n(\tilde x^n),\label{eq:coding2}
\end{align}
where we used the definition of $P'$ and Lemma \ref{lem:AEP} in $(i)$. By Lemma \ref{lem:verschtypen},
\[
  P_{\bar Y_r}^n(\mathcal T_{\bar Y_{r'}\vert\bar X,\delta}^n(\tilde x^n))\leq(n+1)^{\lvert\mathcal A\rvert\lvert\mathcal B\rvert}2^{-n(I(\bar X\wedge\bar Y_{r'})-f_3(\delta))}.
\]
This immediately gives
\[
  \eqref{eq:coding2}\leq(1-2^{-nc'\delta})^{-2}(n+1)^{\lvert\mathcal A\rvert\lvert\mathcal B\rvert}2^{-n(I(\bar X\wedge\bar Y_{r'})-f_3(\delta))},
\]
and we can upper-bound \eqref{eq:coding1} by
\begin{align*}
  \lvert\mathcal R'\rvert J_n L_n\exp\bigl\{-n(\min_{r'\in\mathcal R'}I(\bar X\wedge\bar Y_{r'})-2f_3(\delta))\bigr\}.
\end{align*}
If one chooses $\delta$ so small that $\tau\geq4f_3(\delta)>0$, this tends to $0$ exponentially. Combining the bounds on \eqref{eq:coding3} and \eqref{eq:coding1}, we thus obtain \eqref{eq:errorgivenq} for some appropriate $a'>0$.

Using the Markov inequality and setting $a:=a'/3$, we obtain from \eqref{eq:errorgivenq}
\begin{multline*}
  \mathbb P\Biggl[\bigcap_{r\in\mathcal R'}\left\{e_r(\mathcal K_n^{\mathcal X})\leq2^{-na}\right\}\Biggr]
  \geq 1-\sum_{r\in\mathcal R'}\mathbb P[e_r(\mathcal K_n^{\mathcal X})>2^{-na}]\\
  \geq 1-2^{na}\sum_{r\in\mathcal R'}\mathbb E[e_r(\mathcal K_n^{\mathcal X})]
  \geq 1-\lvert\mathcal R'\rvert 2^{na}2^{-3na}\geq 1-2^{-na}
\end{multline*}
for sufficiently large $n$. Thus the probability that $\max_{r\in\mathcal R'}e_r(\mathcal K_n^{\mathcal X})\leq 2^{-na}$ is lower-bounded by $1-2^{-na}$. This completes the proof.
\end{IEEEproof}

We now appeal to the approximation argument of \cite{BBT}, from which we conclude that the same decoding sets induce an exponentially decreasing average error for the complete state set $\mathcal R$ with the same probability lower-bounded by $1-2^{-na}$. This is still true for a non-stochastic encoder, the randomization index can still be decoded. 

Now recall the definition of $E^{\mathcal X}$. Together with the decoding sets
\[
	\mathcal D_j^{\mathcal X}:=\bigcup_{l\in\mathcal L_n}\tilde D_{jl},
\]
for $j\in\mathcal J_n$, this defines a randomly chosen uncorrelated $(n,J_n)$-code $\mathcal K_n^{\mathcal X}$. Note that
\begin{align*}
	\frac{1}{J_n}\sum_{j\in\mathcal J_n}\sum_{x^n}E^{\mathcal X}(x^n\vert j)W^n((\mathcal D_j^{\mathcal X})^c\vert x^n)
	&=\frac{1}{J_nL_n}\sum_{j\in\mathcal J_n}\sum_{l\in\mathcal L_n}W^n((\mathcal D_j^{\mathcal X})^c\vert X_{jl})\\
	&\leq\frac{1}{J_nL_n}\sum_{j\in\mathcal J_n}\sum_{l\in\mathcal L_n}W^n((\hat{\mathcal D}_{jl}^{\mathcal X})^c\vert X_{jl}).
\end{align*}
This last term is exponentially small with high probability by the previous considerations, which proves  Lemma \ref{lem:transmission}.

\section{Proof of Lemma \ref{lem:TV-sec}}\label{app:TV-sec}

Below we will define events $\iota_1(j,z^n,s^n)$ and $\iota_2(j,s^n)$, for $j\in\mathcal J_n$, $z^n\in\mathcal Z^n$ and $s^n\in\mathcal S^n$, and show that the $\iota_0$ defined in Lemma \ref{lem:TV-sec} satisfies
\begin{equation}\label{eq:sub-intersection}
	\iota_0\supset\bigcap_{j,z^n,s^n}\iota_1(j,z^n,s^n)\cap\bigcap_{j,s^n}\iota_2(j,s^n).
\end{equation}
Then to show that $\mathbb P[\iota_0]>1-2^{-\tau_2n}$, it remains to prove that each of the events of the right-hand side of \eqref{eq:sub-intersection} has a probability sufficiently close to 1.

\subsubsection{Definition of $\iota_1(j,z^n,s^n)$}

For some positive $\alpha$ to be chosen later, let $\varepsilon_n:=2^{-n\alpha}$. Fix $s^n\in\mathcal S^n$, and denote its type by $q\in\mathcal P_0^n(\mathcal S)$. For $x^n\in\mathcal A^n$, define
\[
  \mathcal E_1(x^n,s^n):=\bigl\{z^n\in\mathcal T_{\bar Z_q,4\lvert\mathcal A\rvert\lvert\mathcal S\rvert\delta}^n:V^n_{s^n}(z^n\vert x^n)\leq\exp\{-n(H(\bar Z_q\vert\bar X)-f_2(3\lvert\mathcal S\rvert\delta))\}\bigr\},
\]
where $f_2$ is the function from Lemma \ref{lem:wknf}, and set
\begin{equation}\label{eq:thetatilde}
  \tilde\Theta_{s^n}(z^n):=\mathbb E[V^n_{s^n}(z^n\vert X_{11})\mathbbm 1_{\mathcal E_1(X_{11},s^n)}(z^n)].
\end{equation}
Further define
\[
  \mathcal E_2(s^n):=\bigl\{z^n\in\mathcal T_{\bar Z_q,4\lvert\mathcal A\rvert\lvert\mathcal S\rvert\delta}^n:\tilde\Theta_{s^n}(z^n)\geq\varepsilon_n\lvert\mathcal T_{\bar Z_q,4\lvert\mathcal A\rvert\lvert\mathcal S\rvert\delta}^n\rvert^{-1}\bigr\}
\]
and set
\[
  \Theta_{s^n}(z^n):=\tilde\Theta_{s^n}(z^n)\mathbbm 1_{\mathcal E_2(s^n)}(z^n).
\]
Note that by definition, $\Theta_{s^n}(z^n)>0$ only if $z^n\in\mathcal T_{\bar Z_q,4\lvert\mathcal A\rvert\lvert\mathcal S\rvert\delta}^n$.

With the sets just defined, we obtain a modification of $V_{s^n}^n$ by defining
\[
  Q_{s^n,z^n}(x^n):=V^n_{s^n}(z^n\vert x^n)\mathbbm 1_{\mathcal E_1(x^n,s^n)}(z^n)\mathbbm 1_{\mathcal E_2(s^n)}(z^n).
\]
Note that this is not an actual ``channel'' as in general $\sum_{z^n}Q_{s^n,z^n}(x^n)<1$. Finally, we define
\[
  \iota_1(j,z^n,s^n):=\left\{\frac{1}{L_n}\sum_{l=1}^{L_n}Q_{s^n,z^n}(X_{jl})\in[(1\pm\varepsilon_n)\Theta_{s^n}(z^n)]\right\},
\]
where $[(1\pm\varepsilon_n)\Theta_{s^n}(z^n)]$ is short for $[(1-\varepsilon_n)\Theta_{s^n}(z^n),(1+\varepsilon_n)\Theta_{s^n}(z^n)]$.

\subsubsection{Definition of $\iota_2(j,s^n)$}

Let $q\in\mathcal P_0^n(\mathcal S)$ be the type of $s^n$ and let $\bar S_q$ be an $\mathcal S$-valued random variable independent of the family $\{\bar X,\bar Y_r,\bar Z_q:r\in\mathcal R,q\in\mathcal P(\mathcal S)\}$ with $P_{\bar S_q}=q$. Then we define
\[
	\iota_2(j,s^n):=\left\{\lvert\{l\in\mathcal L_n:s^n\in T_{\bar S_q\vert\bar X,2\delta}^n(X_{jl})\}\rvert\geq(1-\varepsilon_n)(1-2^{-nc'\delta^2})L_n\right\}.
\]

\subsubsection{Proof of Lemma \ref{lem:TV-sec}}

The proof of Lemma \ref{lem:TV-sec} bases on three lemmas. The first one proves that \eqref{eq:sub-intersection} actually is true. 

\begin{lem}\label{lem:thisthenthat}
Assume a realization $\mathbf x:=\{x_{jl}:j\in\mathcal J_n,l\in\mathcal L_n\}$ of $\mathcal X$ has the following properties: For all $j\in\mathcal J_n$ and $z^n\in\mathcal C^n$ and $q\in\mathcal P_0^n(\mathcal S)$ and $s^n\in S^n$,
\begin{align}
  \frac{1}{L_n}\sum_{l=1}^{L_n}Q_{s^n,z^n}(x_{jl})&\in[(1\pm\varepsilon_n)\Theta_{s^n}(z^n)],\label{eq:prop1}\\
  \frac{\lvert\{l\in\mathcal L_n:s^n\in T_{\bar S_q,2\delta}^n(x_{jl})\}\rvert}{L_n}&\geq(1-\varepsilon_n-2^{-nc'\delta^2}),\label{eq:prop2}.
\end{align}
Then 
\[
	\max_{j\in\mathcal J_n}\max_{s^n\in\mathcal S^n}\lVert P_{Z^n_{s^n}\vert M^n}(\,\cdot\,\vert j)-\Theta_{s^n}(\cdot)\lVert\leq4(\varepsilon_n+2^{-nc'\delta^2}).
\]
In particular, \eqref{eq:sub-intersection} is true with $\tau_1=\min\{\alpha,c'\delta^2\}/2$.
\end{lem}

This lemma is proved in Appendix \ref{app:sec-proofs}. The next two lemmas bound the probabilities of the complements of the $\iota_1$ and $\iota_2$ sets.

\begin{lem}\label{lem:modifchannel}
For sufficiently small $\delta>0$ there exists a $\tau_3>0$ such that for $n$ large and every $j\in\mathcal J_n,z^n\in\mathcal C^n$ and $s^n\in\mathcal S^n$ 
\[
  \mathbb P\bigl[\iota_1(j,z^n,s^n)^c\bigr]\leq 2\exp\Bigl\{-\exp\bigl\{n\tau_3\bigr\}\Bigr\}.
\]
\end{lem}

\begin{lem}\label{lem:enoughls}
  For every $j\in\mathcal J_n$ and $s^n\in\mathcal S^n$,
\[
   \mathbb P[\iota_2(j,s^n)]\leq 2\exp\Bigl\{-\exp\bigl\{n(\max_{q\in\mathcal P(\mathcal S)}I(\bar X\wedge\bar Z_q)+\tau_5)\}\Bigr\}
\]
for some $\tau_5>0$, provided that $n$ is sufficiently large.
\end{lem}

The proofs of Lemmas \ref{lem:modifchannel} and \ref{lem:enoughls} can be found in Appendix \ref{app:sec-proofs}. They show that the probability of the complement of each of the events $\iota_1(j,z^n,s^n)$ and $\iota_2(j,s^n)$ is upper-bounded by a term which tends to zero doubly-exponentially as the blocklength increases. Then
\begin{align*}
	\mathbb P[\iota_0]
	&=1-\mathbb P[\iota_0^c]\\
	&\stackrel{(i)}{\geq}1-\mathbb P\left[\bigcup_{j,z^n,s^n}\iota_1(j,z^n,s^n)^c\cup\bigcup_{j,s^n}\iota_2(j,s^n)^c\right]\\
  &\stackrel{(ii)}{\geq} 1-2J_n\lvert\mathcal C\rvert^n\lvert\mathcal S\rvert^n\exp\bigl\{-\exp\{n\tau_3\}\bigr\}-2J_n\lvert\mathcal S\rvert^n\exp\bigl\{-\exp\{n(\max_{q\in\mathcal P(\mathcal S)}I(\bar X\wedge\bar Z_q)+\tau_5)\}\bigr\}\\
  &\stackrel{(iii)}{\geq}1-2^{-n\tau_1},
\end{align*}
where (i) is due to \eqref{eq:sub-intersection}, (ii) to the union bound and (iii) because an appropriate $\tau_1>0$ can be found due to the doubly exponential decrease of the probabilities in Lemmas \ref{lem:modifchannel} and \ref{lem:enoughls}. Altogether, this proves Lemma \ref{lem:TV-sec}.

\section{Proofs of Lemmas \ref{lem:thisthenthat}-\ref{lem:enoughls}}\label{app:sec-proofs}

\subsection{Proof of Lemma \ref{lem:modifchannel}}\label{subsect:chernoff}

Let $j\in\mathcal J_n,z^n\in\mathcal C^n,s^n\in\mathcal S^n$. We want to upper-bound the probability of the event that
\[
	\left\{\frac{1}{L_n}\sum_{l=1}^{L_n}Q_{s^n,z^n}(X_{jl})\notin[(1\pm\varepsilon_n)\Theta_{s^n}(z^n)]\right\}.
\]
The form of this event already suggests that a Chernoff bound may be the right method for the proof. Indeed, we will apply the following lemma.

\begin{lem}\label{lem:Chernoff}
  Let $b$ be a positive number. Let $Z_1,\ldots,Z_L$ be i.i.d.\ random variables with values in $[0,b]$ and expectation $\mathbb EZ_l=\nu$, and let $0<\varepsilon<\frac{1}{2}$. Then
\[
  \mathbb P\left\{\frac{1}{L}\sum_{l=1}^LZ_i\notin[(1\pm\varepsilon)\nu]\right\}\leq 2\exp\left(-L\cdot\frac{\varepsilon^2\nu}{3b}\right).
\]
\end{lem}

\begin{IEEEproof}
  The proof can be found in \cite[Theorem 1.1]{DP} and in \cite{AW}.
\end{IEEEproof}

The claim of Lemma \ref{lem:modifchannel} follows from an application of Lemma \ref{lem:Chernoff}. Due to the definition of $\mathcal E_1(x^n,s^n)$, the random variables $Q_{s^n,z^n}(X_{jl})$ are upper-bounded by $\exp\{-n(H(\bar Z_q\vert\bar X)-f_2(3\lvert\mathcal S\rvert\delta))\}$ and have mean $\Theta_{s^n}(z^n)$. Lemma \ref{lem:wknf} implies that $\Theta_{s^n}(z^n)\geq\varepsilon_n\exp\{-n(H(\bar Z_q)+f_1(4\lvert\mathcal A\rvert\lvert\mathcal S\rvert\delta))\}$. Inserting this into the right-hand side of Lemma \ref{lem:Chernoff} and recalling the definition of $\varepsilon_n$ gives the desired bound, with $\tau_3=\tau/5-3\alpha-f_1(4\lvert\mathcal A\rvert\lvert\mathcal S\rvert\delta)-f_2(3\lvert\mathcal S\rvert\delta)$. This is positive if $\alpha$ and $\delta$ are sufficiently small. This proves Lemma \ref{lem:modifchannel}.

\subsection{Proof of Lemma \ref{lem:enoughls}}\label{subsect:goodcodewords}

The proof also applies the Chernoff bound of Lemma \ref{lem:Chernoff}. To do so, we need a lower bound on $\mathbb E[\mathbbm 1_{\mathcal T^n_{\bar S_q,2\delta}(X_{11})}]=\mathbb P[s^n\in\mathcal T_{\bar S_q,2\delta}^n(X_{11})]$.

\begin{lem}\label{lem:typesubset}
  For sufficiently large $n$ and every $s^n$ of type $q$,
\[
  \mathbb P[s^n\in\mathcal T_{\bar S_q,2\delta}^n(X_{11})]\geq 1-2^{-nc'\delta^2}.
\]
\end{lem}

\begin{IEEEproof}
  We first show
\begin{align}
  \mathcal T_{\bar X\vert\bar S_q,\delta/\lvert S\rvert}^n(s^n)\subset\{x^n\in\mathcal T_{\bar X,\delta}^n:s^n\in\mathcal T_{\bar S_q,2\delta}^n(x^n)\}.\label{eq:SAAS2}
\end{align}
Let $x^n\in\mathcal T_{\bar X\vert\bar S_q,\delta/\lvert\mathcal S\rvert}^n(s^n)$. Clearly $\mathcal T_{\bar X\vert\bar S_q,\delta/\lvert\mathcal S\rvert}^n(s^n)\subset\mathcal T_{\bar X,\delta}^n$. Then
\begin{align*}
  &\hphantom{\mathrel{=}}\;\,\left\lvert\frac{1}{n}N(s,a\vert s^n,x^n)-P_{\bar S_q\vert\bar X}(s\vert a)\frac{1}{n}N(a\vert x^n)\right\rvert\\
  &=\left\lvert\frac{1}{n}N(s,a\vert s^n,x^n)-\frac{1}{n}N(s\vert s^n)\frac{1}{n}N(a\vert x^n)\right\rvert\\
  &\leq\left\lvert\frac{1}{n}N(s,a\vert s^n,x^n)-P_{\bar X\vert\bar S_q}(a\vert s)\frac{1}{n}N(s\vert s^n)\right\rvert\\
  &\quad+\frac{1}{n}N(s\vert s^n)\left\lvert P_{\bar X}(a)-\frac{1}{n}N(a\vert x^n)\right\rvert\\
  &\leq\frac{\delta}{\lvert\mathcal S\rvert}+\delta\leq2\delta.
\end{align*}
This proves \eqref{eq:SAAS2}. For $n$ large, we can use this to continue with
\begin{align*}
  &\mathbb P[s^n\in\mathcal T_{\bar S_q\vert\bar X,2\delta}^n(X_{11})]
  \stackrel{(i)}{\geq}\mathbb P[\mathcal T_{\bar X\vert\bar S_q,\delta/\lvert\mathcal S\rvert}^n(s^n)]
  =\sum_{x^n\in\mathcal T_{\bar X\vert\bar S_q,\delta/\lvert\mathcal S\rvert}^n(s^n)}p'(x^n)\\
  &\stackrel{(ii)}{\geq}\sum_{x^n\in\mathcal T_{\bar X\vert\bar S_q,\delta/\lvert\mathcal S\rvert}^n(s^n)}P_{\bar X}^n(x^n)\\
  &=P_{\bar X\vert\bar S_q}^n(\mathcal T_{\bar X\vert\bar S_q,\delta/\lvert\mathcal S\rvert}^n(s^n)\vert s^n)\\
  &\stackrel{(iii)}{\geq}1-2^{-nc'\delta^2},
\end{align*}
where we used \eqref{eq:SAAS2} in $(i)$, $\mathcal T_{\bar X\vert\bar S_q,\delta/\lvert\mathcal S\rvert}^n(s^n)\subset\mathcal T_{\bar X,\delta}^n$ in $(ii)$ and Lemma \ref{lem:AEP} in $(iii)$.
\end{IEEEproof}

Moving to the proof of Lemma \ref{lem:enoughls}, let $j\in\mathcal J_n$. The i.i.d.\ random variables $\mathbbm 1_{T_{\bar S_q\vert\bar X,2\delta}^n(X_{jl})}(s^n)$ ($l\in\mathcal L_n$) are upper-bounded by $1$. Their expectation $\nu$ was lower-bounded in Lemma \ref{lem:typesubset}  by $1-2^{-nc'\delta^2}$. This implies that $\iota_2(j,s^n)^c$ is contained in the event
\[
  \left\{\frac{1}{L_n}\lvert\{l\in\mathcal L_n:s^n\in T_{\bar S_q\vert\bar X,2\delta}^n(X_{jl})\}\rvert\leq(1-\varepsilon_n)\nu \right\}.
\]
Lemma \ref{lem:Chernoff} thus implies that the probability of the above event is upper-bounded as claimed if $n$ is large enough upon setting $\tau_5:=\tau/4-3\alpha$ and letting $\alpha$ be small enough.

\subsection{Proof of Lemma \ref{lem:thisthenthat}}

The next two lemmas are needed for the proof. Recall the convention that we sometimes write $V(c\vert a,s)$ instead of $V_s(c\vert a)$. 

\begin{lem}\label{lem:whatfor}
  Let $x^n\in\mathcal T_{\bar X,\delta}^n$ and let $s^n$ have type $q\in\mathcal P_0^n(\mathcal S)$. Let the random variable $\underline Z_q$ satisfy $P_{\underline Z_q\vert\bar X\bar S_q}(\cdot\vert\cdot,\cdot)=V(\cdot\vert\cdot,\cdot)$. If $s^n\in\mathcal T_{\bar S_q,2\delta}^n(x^n)$, then $\mathcal T_{\underline Z_q\vert\bar X\bar S_q,\delta}^n(x^n,s^n)\subset \mathcal E_1(x^n,s^n)$.
\end{lem}

\begin{IEEEproof}
  For $x^n\in\mathcal T_{\bar X,\delta}^n$, we have $\mathcal T_{\bar Z_q\vert\bar X,3\lvert\mathcal S\rvert\delta}^n(x^n)\subset\mathcal T_{\bar Z_q,4\lvert\mathcal A\rvert\lvert\mathcal S\rvert\delta}^n$. Thus due to Lemma \ref{lem:wknf}, it suffices to show that if $s^n$ has type $q$, then $\mathcal T_{\underline Z_q\vert\bar X\bar S_q,\delta}^n(x^n,s^n)\subset\mathcal T_{\bar Z_q\vert\bar X,3\lvert\mathcal S\rvert\delta}^n(x^n)$. For $a\in\mathcal A$ and $c\in\mathcal C$, we calculate
\begin{align*}
  &\left\lvert\frac{1}{n}N(c,a\vert z^n,x^n)-\sum_{s\in\mathcal S}q(s)V(c\vert a,s)\frac{1}{n}N(a\vert x^n)\right\rvert\\
  &\leq\sum_{s\in\mathcal S}\left\lvert\frac{1}{n}N(c,a,s\vert z^n,x^n,s^n)-q(s)V(c\vert a,s)\frac{1}{n}N(a\vert x^n)\right\rvert\\
  &\leq\sum_{s\in\mathcal  S}\left\lvert\frac{1}{n}N(c,a,s\vert z^n,x^n,s^n)-V(c\vert a,s)\frac{1}{n}N(a,s\vert x^n,s^n)\right\rvert\\
  &+\sum_{s\in\mathcal S}V(c\vert a,s)\left\lvert\frac{1}{n}N(a,s\vert x^n,s^n)-q(s)\frac{1}{n}N(a\vert x^n)\right\rvert\\
  &\leq\lvert\mathcal S\rvert(\delta+2\delta)=3\lvert\mathcal S\rvert\delta. 
\end{align*}
\end{IEEEproof}

\begin{cor}\label{cor:untabsch}
  If $n$ is sufficiently large, then every $s^n\in\mathcal S^n$ satisfies
\[
  \Theta_{s^n}(\mathcal C^n)\geq 1-2\cdot2^{-nc'\delta^2}-\varepsilon_n
\]
\end{cor}

\begin{IEEEproof}
  Let $s^n$ have type $q\in\mathcal P_0^n(\mathcal S)$. By the definition of $\Theta_{s^n}$, we have $\Theta_{s^n}(\mathcal C^n)=\Theta_{s^n}(\mathcal E_2(s^n))$. As the support of $\tilde\Theta_{s^n}$ is contained in $T_{\bar Z_q,4\lvert\mathcal A\rvert\lvert\mathcal S\rvert\delta}^n$, we have $\Theta_{s^n}(\mathcal E_2(s^n))\geq\tilde\Theta_{s^n}(\mathcal T_{\bar Z_q,4\lvert\mathcal A\rvert\lvert\mathcal S\rvert\delta}^n)-\varepsilon_n=\tilde\Theta_{s^n}(\mathcal C^n)-\varepsilon_n$. By definition,
\begin{align*}
  \tilde\Theta_{s^n}(\mathcal C^n)&=\mathbb E[V^n_{s^n}(\mathcal E_1(X_{11},s^n)\vert X_{11})]\\
  &\geq\mathbb E[V^n_{s^n}(\mathcal E_1(X_{11},s^n)\vert X_{11})\vert s^n\in\mathcal T_{\bar S_q\vert\bar X,2\delta}^n(X_{11})]\mathbb P[s^n\in\mathcal T_{\bar S_q\vert\bar X,2\delta}^n(X_{11})].
\end{align*}
For sufficiently large $n$
\begin{align*}
  &\mathbb E[V^n_{s^n}(\mathcal E_1(X_{11},s^n)\vert X_{11})\vert s^n\in\mathcal T_{\bar S_q\vert\bar X,2\delta}^n(X_{11})]\\
  &\stackrel{(i)}{\geq}\mathbb E[V^n(\mathcal T_{\underline Z_q\vert\bar X\bar S_q,\delta}^n(X_{11},s^n)\vert X_{11},s^n)\vert s^n\in\mathcal T_{\bar S_q\vert\bar X,2\delta}^n(X_{11})]\\
  &\stackrel{(ii)}{\geq} 1-2^{-nc'\delta^2},
\end{align*}
where we used Lemma \ref{lem:whatfor} in $(i)$ and Lemma \ref{lem:AEP} in $(ii)$. Lemma \ref{lem:typesubset} provides a lower bound on $\mathbb P[s^n\in\mathcal T_{\bar S_q\vert\bar X,2\delta}^n(X_{11})]$, so altogether,
\begin{equation}\label{eq:Theta}
  \Theta_{s^n}(\mathcal C^n)\geq\tilde\Theta_{s^n}(\mathcal C^n)-\varepsilon_n\geq(1-2^{-nc'\delta^2})^2-\varepsilon_n\geq1-2\cdot2^{-nc'\delta^2}-\varepsilon_n. 
\end{equation}
\end{IEEEproof}

	Let $\mathbf x=\{x_{jl}:j\in\mathcal J_n,l\in\mathcal L_n\}$ be a realization of $\mathcal X$ satisfying \eqref{eq:prop1} and \eqref{eq:prop2}. Let $\mathcal K_n$ be the corresponding code and $\mathcal F(\mathcal K_n,\mathfrak{\overline W,V})=\{M^n,X^n,Y_r^n,Z_{s^n}^n,\hat M_r:r\in\mathcal R,s^n\in\mathcal S^n\}$ the canonical family of random variables associated with $\mathcal K_n$. For any $s^n$ with type $q\in\mathcal P_0(\mathcal S)$, we decompose the total variation distance as follows:
\begin{align}
  &\lVert P_{Z^n_{s^n}\vert M^n}(\,\cdot\,\vert j)-\Theta_{s^n}(\cdot)\lVert\notag\\
  &\leq\left\lVert\frac{1}{L_n}\sum_{l=1}^{L_n}Q_{s^n,\,\cdot\,}(x_{jl})-\Theta_{s^n}(\cdot)\right\rVert\label{eq:erster}\\
  &+\left\lVert\frac{1}{L_n}\sum_{l=1}^{L_n}V^n_{s^n}(\,\cdot\,\vert x_{jl})\mathbbm 1_{\mathcal E_1(x_{jl},s^n)}(\cdot)(\mathbbm 1_{\mathcal C^n}(\cdot)-\mathbbm 1_{\mathcal E_2(s^n)}(\cdot))\right\rVert\label{eq:zweiter}\\
  &+\left\lVert\frac{1}{L_n}\sum_{l=1}^{L_n}V^n_{s^n}(\,\cdot\,\vert x_{jl})(\mathbbm 1_{\mathcal C^n}(\cdot)-\mathbbm 1_{\mathcal E_1(x_{jl},s^n)}(\cdot))\right\rVert.\label{eq:dritter}
\end{align}

The term in \eqref{eq:erster} is upper-bounded by $\varepsilon_n$, because due to \eqref{eq:prop1}
\begin{align*}
  &\left\lVert\frac{1}{L_n}\sum_{l=1}^{L_n}Q_{s^n,\,\cdot\,}(x_{jl})-\Theta_{s^n}(\cdot)\right\rVert\\
  &=\sum_{z^n}\left\lvert\frac{1}{L_n}\sum_{l=1}^{L_n}Q_{s^n,z^n}(x_{jl})-\Theta_{s^n}(z^n)\right\rvert\\
  &\leq\varepsilon_n\sum_{z^n}\Theta_{s^n}(z^n)\\
  &\leq\varepsilon_n.
\end{align*}

Next, applying \eqref{eq:prop1} in $(i)$, we upper-bound \eqref{eq:zweiter} as
\begin{align*}
  &\frac{1}{L_n}\sum_{l=1}^{L_n}\sum_{z^n}V_{s^n}(z^n\vert x_{jl})\mathbbm 1_{\mathcal E_1(x_{jl},s^n)}(z^n)\\
  &-\frac{1}{L_n}\sum_{l=1}^{L_n}\sum_{z^n}V_{s^n}(z^n\vert x_{jl})\mathbbm 1_{\mathcal E_1(x_{jl},s^n)}(z^n)\mathbbm 1_{\mathcal E_2(s^n)}(z^n)\\
  &\leq 1-\sum_{z^n}\frac{1}{L_n}\sum_{l=1}^{L_n}Q_{s^n,z^n}(x_{jl})\\
  &\stackrel{(i)}{\leq}1-(1-\varepsilon_n)\Theta_{s^n}(\mathcal C^n).
\end{align*}
Upon application of Corollary \ref{cor:untabsch}, we obtain that \eqref{eq:zweiter} can be upper-bounded by
\[
  1-(1-\varepsilon_n)(1-2\cdot2^{-nc'\delta^2}-\varepsilon_n)\leq2(2^{-nc'\delta}+\varepsilon_n).
\]

It remains to upper-bound \eqref{eq:dritter}. Recall the definition of $\underline Z_q$. We have
\begin{align}\label{eq:ubIII}
  \left\lVert\frac{1}{L_n}\sum_{l=1}^{L_n}V^n_{s^n}(\,\cdot\,\vert x_{jl})(\mathbbm 1_{\mathcal C^n}(\cdot)-\mathbbm 1_{\mathcal E_1(x_{jl},s^n)}(\cdot))\right\rVert\notag\\
  =\frac{1}{L_n}\sum_{l=1}^{L_n} V^n_{s^n}(\mathcal E_1(x_{jl},s^n)^c\vert x_{jl})\\
  =\frac{1}{L_n}\sum_{\substack{l\in\mathcal L_n:\\\mathcal T_{\underline Z_q\vert\bar X\bar S_q,\delta}^n(x_{jl},s^n)\subset \mathcal E_1(x_{jl},s^n)}} V^n_{s^n}(\mathcal E_1(x_{jl},s^n)^c\vert x_{jl})\notag\\
  +\frac{1}{L_n}\sum_{\substack{l\in\mathcal L_n:\\\mathcal T_{\underline Z_q\vert\bar X\bar S_q,\delta}^n(x_{jl},s^n)\nsubseteq \mathcal E_1(x_{jl},s^n)}} V^n_{s^n}(\mathcal E_1(x_{jl},s^n)^c\vert x_{jl}).\notag
\end{align}
If $\mathcal T_{\underline Z_q\vert\bar X\bar S_q,\delta}^n(x_{jl},s^n)\subset \mathcal E_1(x_{jl},s^n)$, then by Lemma \ref{lem:AEP}, we have
\[
  V^n_{s^n}(\mathcal E_1(x_{jl},s^n)^c\vert x_{jl})\leq V^n(\mathcal T_{\underline Z_q\vert\bar X\bar S_q,\delta}^n(x_{jl},s^n)^c\vert x_{jl},s^n)\leq2^{-nc'\delta^2}.
\]
By Lemma \ref{lem:whatfor} and \eqref{eq:prop2}, the proportion of those $j$ for which $\mathcal T_{\underline Z_q\vert\bar X\bar S_q,\delta}^n(x_{jl},s^n)\nsubseteq \mathcal E_1(x_{jl},s^n)$ holds is upper-bounded by $\varepsilon_n+2^{-nc'\delta^2}$. We can thus bound \eqref{eq:ubIII} by
\[
  2^{-nc'\delta^2}+\varepsilon_n+2^{-nc'\delta}=\varepsilon_n+2\cdot2^{-nc'\delta^2}.
\]

Collecting the bounds on \eqref{eq:erster}, \eqref{eq:zweiter} and \eqref{eq:dritter} completes the proof of Lemma \ref{lem:thisthenthat}.

\end{document}